\documentclass{aa} % 
\usepackage{times,graphicx,epsfig,color}
\usepackage[colorlinks=true,allcolors=blue]{hyperref}
\usepackage{natbib}
\begin{document}

\title{Local disc model in view of Gaia DR1 and RAVE data}

\author{K. Sysoliatina\inst{1}\thanks{Fellow of the International Max Planck Research School for Astronomy
and Cosmic Physics at the University of Heidelberg (IMPRS-HD).}, 
A. Just\inst{1},
I. Koutsouridou\inst{2}, 
E.K. Grebel\inst{1}, 
G. Kordopatis\inst{8},
M. Steinmetz\inst{10},
O. Bienaym\'{e}\inst{9},
B.K. Gibson\inst{3}, 
J. Navarro\inst{7},
W. Reid\inst{5,6}, \and
G. Seabroke\inst{4}
}

%\offprints{K. Sysoliatina}
\mail{Sysoliatina@uni-heidelberg.de}

\institute{
$^{1}$Astronomisches Rechen-Institut, Zentrum f\"{u}r Astronomie der Universit\"{a}t Heidelberg, M\"{o}nchhofstr. 12--14, 69120 Heidelberg, Germany\\
$^{2}$Observatoire de Paris, GEPI, 77 av. Denfert-Rocherau, 75014 Paris, France\\
$^{3}$E.A. Milne Centre for Astrophysics, University of Hull, Hull, HU6 7RX, UK\\
$^{4}$Mullard Space Science Laboratory, University College London, Holmbury St Mary, Dorking, RH5 6NT, UK\\
$^{5}$Department of Physics and Astronomy, Macquarie University, Sydney, NSW 2109, Australia\\
$^{6}$Western Sydney University, Locked bag 1797, Penrith South DC, NSW 2751, Australia\\
$^{7}$Senior CIfAR Fellow, University of Victoria, Victoria BC, Canada V8P 5C2\\
$^{8}$Universit\'{e} C\^{o}te d'Azur, Observatoire de la C\^{o}te d'Azur, CNRS, Laboratoire Lagrange, France\\
$^{9}$Observatoire astronomique de Strasbourg, Universit\'{e} de Strasbourg, CNRS, 11 rue de l'Universit\'{e}, F-67000 Strasbourg, France\\
$^{10}$Leibniz Institut f\"{u}r Astrophysik Potsdam, An der Sternwarte 16, D-14482, Potsdam, Germany 
}

\date{Printed: \today}

\abstract 
%Context
{}
%Aims
{We test the performance of the semi-analytic self-consistent Just-Jahrei{\ss} disc model (JJ model) with
the astrometric data from the Tycho-Gaia Astrometric Solution (TGAS) sub-catalogue 
of the first Gaia data release (Gaia DR1), as well as the radial velocities from the fifth data release of the Radial Velocity Experiment survey (RAVE DR5).} 
%Methods
{We used a sample of 19,746 thin-disc stars from the TGAS$\times$RAVE cross-match selected in a local solar cylinder
of 300 pc radius and 1 kpc height below the Galactic plane.
Based on the JJ model, we simulated this sample via the forward modelling technique.
First, we converted the predicted vertical density laws of the thin-disc populations into a mock sample.
For this we used of the Modules and Experiments in Stellar Astrophysics (MESA) Isochrones and Stellar Tracks (MIST),
a star formation rate (SFR) that decreased after a peak at 10 Gyr ago, and a three-slope broken power-law initial mass function (IMF). 
Then the obtained mock populations were reddened with a 3D dust map and were subjected to the selection criteria corresponding to the RAVE and TGAS observational 
limitations as well as to additional cuts applied to the data sample.
We calculated the quantities of interest separately at different heights above the Galactic plane, taking into account the distance error effects
in horizontal and vertical directions into account separately.}
%Results 
{The simulated vertical number density profile agrees well with the data. 
An underestimation of the stellar numbers begins at $\sim$800 pc from the Galactic plane, which is expected 
as the possible influence of populations from $|z|>1$ kpc is ignored during the modelling.  
The lower main sequence (LMS) is found to be thinner and under-populated by 3.6\% relative to the observations. 
The corresponding deficits for the upper main sequence (UMS) and red giant branch (RGB) are 6\% and 34.7\%, respectively. 
However, the intrinsic uncertainty related to the choice of stellar isochrones is $\sim$10\% in the total stellar number. 
The vertical velocity distribution function f($|W|$) simulated for the whole cylinder agrees to within 1$\sigma$ with the data. 
This marginal agreement arises because the dynamically cold populations at heights $<200$ pc from the Galactic plane are underestimated.  
We find that the model gives a fully realistic representation of the vertical gradient in stellar populations when 
studying the Hess diagrams for different horizontal slices. We also checked and confirm the 
consistency of our results with the newly available second Gaia data release (DR2).}
%Conclusions 
{Based on these results and considering the uncertainties in the data selection as well as the sensitivity of the simulations to the 
sample selection function, we conclude that the fiducial JJ model confidently reproduces the vertical trends in the thin-disc stellar population properties. 
Thus, it can serve as a starting point for the future extension of the JJ model to other Galactocentric distances.}

\keywords{Galaxy: disc -- Galaxy: kinematics and dynamics -- Galaxy: solar neighbourhood -- Galaxy: evolution}

\authorrunning{K. Sysoliatina et al.}
\titlerunning{The local disc model in view of Gaia DR1 and RAVE data}

\maketitle 

%\parskip=0.pt
%%%%%%%%%%%%%%%%%%%%%%%%%%%%%%%%%%%%%%%%%%%%%%%%%%%%%%%%%%%%%%%%%%%%%
%%%%%%%%%%%%%%%%%%%%%%%%%%%%%%%%%%%%%%%%%%%%%%%%%%%%%%%%%%%%%%%%%%%%%
\defcitealias{just10}{Paper~I}
\defcitealias{just11}{Paper~II}
\defcitealias{rybizki15}{Paper~III}

\section{Introduction}\label{intro}

In the past two decades, the amount of data collected on the Milky Way's stellar content has increased by several orders of magnitude. 
A typical present-day large-scale survey contains measurements for millions of objects, and with the European astrometric mission Gaia 
\citep{gaia16,gaia16a}, this number has already increased by a factor of ten. 
The core of the best dataset available so far for the Galactic studies includes photometry from 
the 2 Micron All Sky Survey (2MASS, \mbox{\citealp{skrutskie06}}) and Gaia G-band \citep{jordi10,carrasco16,leeuwen17}, proper motions of 
the PPM-Extended catalogue (PPMX, \mbox{\citealp{roser08}}) and the fifth US Naval Observatory CCD Astrograph Catalog (UCAC5,  \mbox{\citealp{zacharias17}}),
as well as astrometric parameters of the Tycho-Gaia Astrometric Solution (TGAS, \citealp{michalik15,lindegren16}).
Extensive information on the stellar chemical abundances is now available from spectroscopic surveys such as 
GALactic Archaeology with HERMES (GALAH, \mbox{\citealp{martell17}}), 
the Large sky Area Multi-Object Spectroscopic Telescope (LAMOST) General Survey \mbox{\citep{luo15}}, the Sloan Digital Sky Survey III Apache Point Observatory Galactic Evolution Experiment 
(SDSS-III/APOGEE, \mbox{\citealp{eisenstein11}}), 
the Radial Velocity Experiment (RAVE, \mbox{\citealp{steinmetz06}}), and the Gaia-ESO survey \mbox{\citep{gilmore12}}. 
Thus, we are entering an era in which observational data available for Galactic stellar populations 
will be not only various and abundant, but also precise enough to enable the detailed study of the Galactic components and to unravel their evolution. 
To achieve these goals, developed and robust methods are required for a comprehensive analysis of observational data.

% see http://sf2a.eu/proceedings/2017/2017sf2a.conf..0301R.pdf for low V_sun
The most sophisticated current machinery for modelling the Galaxy is the Besan\c con Galaxy model (BGM, \citealp{robin3,robin12,robin17,czekaj14}), 
which accounts for the stellar, gaseous, and non-baryonic content of the Galaxy. It includes the non-axisymmetry of the disc in the form of the bar and the spiral arms
and the disc warp. 
Its first version, presented in \mbox{\citet{robin3}}, is implemented in the tool \textit{Galaxia}\footnote{\url{http://galaxia.sourceforge.net}}
\mbox{\citep{sharma11}}, which allows constructing synthetic catalogues of the Milky Way and accounting for the selection effects of the different surveys. 
Another example of a semi-analytic model is the TRIdimensional modeL of thE GALaxy (TRILEGAL, \mbox{\citealp{girardi05,girardi16}}), which has been designed specifically for star count simulations.   

Each Galactic model has its own focus. In previous papers we developed and presented the semi-analytic Just-Jahrei{\ss} model (JJ model).
Its main purpose is to enable a detailed study of the thin-disc vertical structure (\mbox{\citealp{just10}}, hereafter \mbox{\citetalias{just10}}). 
The JJ model describes an axisymmetric thin disc in a steady state consisting of a set of isothermal stellar populations moving in the total gravitational potential. 
Thick and gaseous discs as well as a dark matter (DM) component are added to the general local mass budget in order to treat the pair of density-potential fully
self-consistently. The main input functions describing the thin-disc evolution are the star formation rate (SFR) and 
age-velocity dispersion relation (AVR), as well as the age-metallicity relation (AMR) and the initial mass function (IMF). 
In \mbox{\citetalias{just10}} the parameters of the JJ model were calibrated with the local kinematics of the main-sequence (MS) stars from the \textit{Hipparcos} 
catalogue \mbox{\citep{vanleeuwen07}}. 
The model was later tested with the SDSS star counts towards the north Galactic pole, 
and the parameters of the thick disc were derived (\mbox{\citealp{just11}}, \mbox{\citetalias{just11}}).
The model IMF was constrained with the sample of \textit{Hipparcos} data combined with the Catalog of Nearby Stars (\mbox{\citealp{rybizki15}}, \mbox{\citetalias{rybizki15}}). 
As discussed in \mbox{\citet{gao13}}, the median model-to-data deviations over the Hess diagrams constructed with the 
SDSS apparent magnitudes and colours towards the north Galactic pole are 5.6\% for the JJ model, but 26\% for TRILEGAL and 20-53\% for the old Besan\c con model \mbox{\citep{robin3}}. 

In the long term, the development of the JJ model is focused on its extension to Galactocentric distances beyond solar. 
A special tool for testing chemical evolution scenarios of the Galaxy, the code Chempy\footnote{\url{https://github.com/jan-rybizki/Chempy}}, has recently been 
developed \mbox{\citep{rybizki17}}.
Constraining Chempy parameters with high-quality spectroscopic data and combining it with the radius-dependent SFR and AVR 
(first steps to the radial extension of these functions are presented in \mbox{\citealp{just16}} and \mbox{\citealp{just18}}) will be 
the road to build a global Milky Way disc model.

In this paper we test the JJ model locally with the high-quality 
astrometric data from the TGAS sub-catalogue of the first Gaia data release (DR1) 
and the radial velocities from the fifth data release of the RAVE survey (DR5, \mbox{\citealp{kunder17}}). 
As more than one billion stars with high-precision astrometry of the second data release (DR2) of Gaia 
have recently become available (published on 25 April, 2018; \citealp{gaia18}), the impact of Gaia DR2 on our results is also discussed.

In the framework of the JJ model, we construct the thin-disc stellar populations in the local solar cylinder, including the full stellar evolution. 
For the sample of mock stellar populations we predict parallaxes as observed in the TGAS$\times$RAVE cross-match, 
although for the purposes of illustration, the results are presented as a function of a simple distance estimate calculated both 
in the data and model as the inverse observed parallax. We also account for the reddening with a realistic 
three-dimensional (3D) extinction model based on the map from \mbox{\citet{green15}} and 
simulate the complicated selection effects. 

This paper has the following structure. 
In Section \ref{data} we describe the TGAS$\times$RAVE data sample and discuss our selection criteria together with the selection functions of the TGAS and RAVE catalogues. 
Section \ref{model} explains the basics of the JJ model and the details of our modelling method. 
Section \ref{results} presents the results of this study. In Section \ref{discus} we discuss the intrinsic uncertainty of the modelling associated 
with the choice of stellar library, and also the usefulness of the applied reddening map and the robustness of our results in view of more
accurate and abundant Gaia DR2 data. Finally, we conclude in Section \ref{final}.

\section{Data}\label{data}

\subsection{TGAS$\times$RAVE thin-disc sample}\label{cuts}

To study kinematics and spatial distribution of the thin-disc stellar populations, the full 6D information in the dynamic phase space  
has to be known for individual stars. The most recent and accurate five astrometric parameters, 
that is, positions, proper motions, and parallaxes, are provided in the TGAS catalogue. 
The astrometric solution was found by combining Gaia measurements of the first 14 months of the mission and 
information on positions from the \textit{Tycho-2} Catalogue \mbox{\citep{hog00}}. 
Typical errors for the parallaxes and proper motions are $\sim$0.3 mas and $\sim$0.3 mas yr$^{-1}$, respectively.  
TGAS covers the whole sky and contains $\sim$2 million stars. 
However, this data sample alone cannot serve for our purposes as it lacks the radial velocities, which prevents deriving 3D stellar space velocity vectors. 
This motivated us to complement TGAS with the information from RAVE DR5 \mbox{\citep{kunder17}}, where not only accurate radial velocities are provided (typical errors
are 1-2 km s$^{-1}$), but also chemical abundances of six elements as well as photometry from other surveys such as 2MASS and 
The AAVSO Photometric All-Sky Survey (APASS, \mbox{\citealp{henden09,munari14}}). 
%RAVE survey covers only southern hemisphere.
The total cross-match between TGAS and RAVE contains 257,288 stars\footnote{The cross-match between catalogues was performed by the RAVE Collaboration, 
see \url{https://www.rave-survey.org/downloads}}. 
For these stars we calculated the estimate of heliocentric distance by inverting the TGAS parallaxes, $\tilde{d}=1/\varpi$. 
Here and elsewhere in this paper notations with tilde are used to emphasise that a given quantity is related to the observed distance derived through this simple inversion. 
We also transformed it to the distance from the Galactic plane $\tilde{z}$.  
Using the TGAS proper motions and parallaxes as well as the RAVE radial velocities, we calculated for each star a 3D velocity in Cartesian coordinates $(U,V,W)$.
The components of the peculiar velocity of the Sun were chosen as $U_\odot=11.1$ \mbox{km s$^{-1}$}, $W_\odot=7.25$ \mbox{km s$^{-1}$} \mbox{\citep{schoenrich10}}, 
$V_\odot=4.47$ \mbox{km s$^{-1}$} \mbox{\citep{sysoliatina18}}. Currently , only the vertical motion of stars is included in the JJ model, 
such that only the $W$ component of the spatial velocities is used in practice. We also calculated the observational errors $(\Delta U,\Delta V, \Delta W)$ 
with the help of the TGAS error covariance matrix. 
\begin{figure}
\centerline{\resizebox{\hsize}{!}{\includegraphics{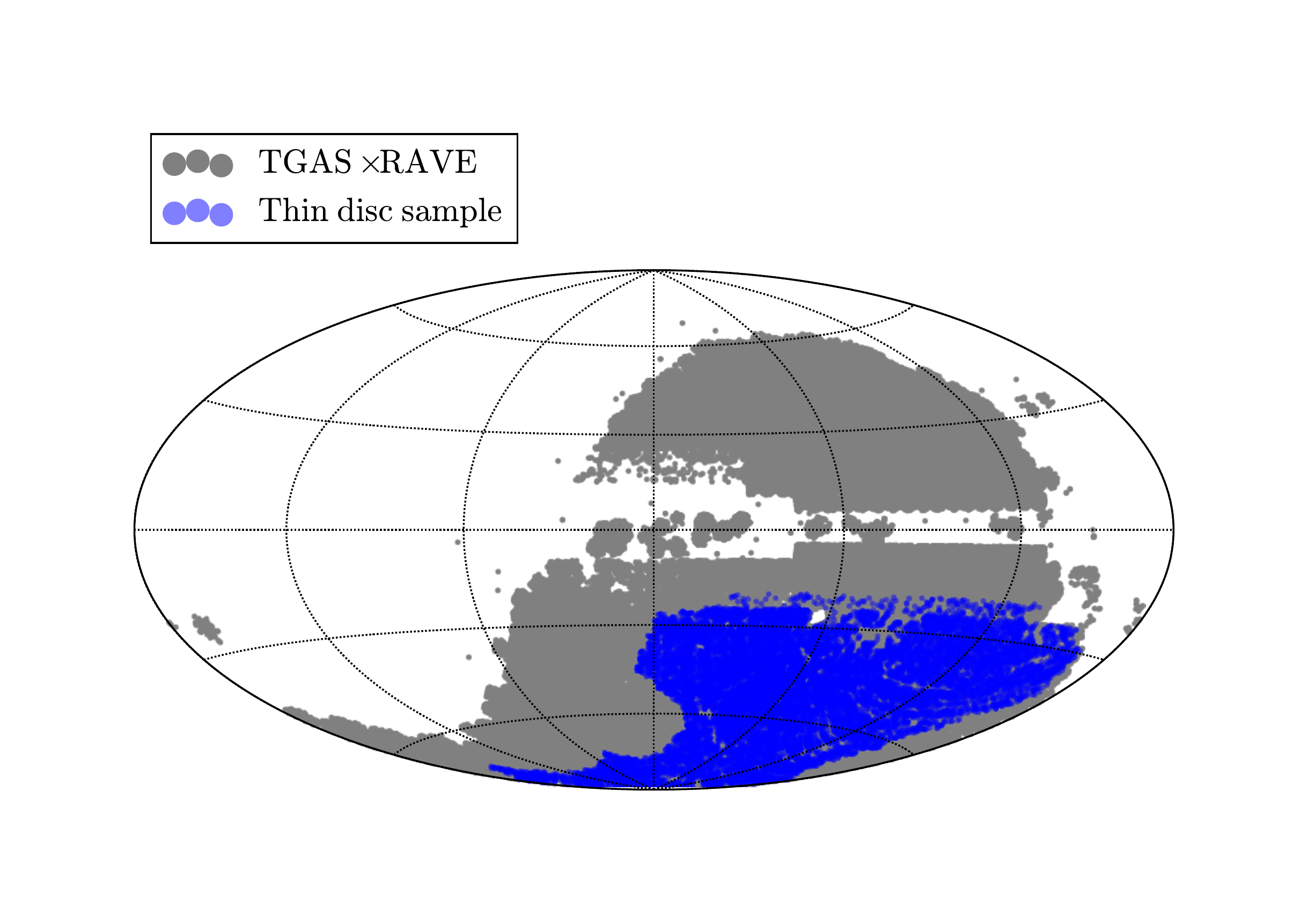}}}
\caption{
Sky coverage of the data in the Galactic coordinates. The full TGAS$\times$RAVE cross-match is shown in grey. 
The selected thin-disc sample (blue) lies below the Galactic plane; its special shape at $b < -50^\circ$ 
reflects the Gaia scanning law pattern. 
}
\label{data_sample}
\end{figure}

The present-day version of the JJ model contains a detailed recipe for constructing the individual populations of the Galactic thin disc while the other components 
are added in a simple way. Thus, to perform a detailed model-to-data comparison, we needed 
to construct a clean thin-disc sample from the data. To do so, we applied the following selection criteria to the TGAS$\times$RAVE cross-match: 
\begin{enumerate}
\item \label{geomcut}\textit{Geometry cut.} 
 We selected stars with $|b|>20^\circ$ as the RAVE survey avoids the Galactic plane and the number of observed stars drops quickly at low Galactic latitudes. 
 %Moreover, the target selection of the input catalogue for the RAVE DR5 changes at latitudes $5^\circ<|b|<25^\circ$ \mbox{\citep{kordopatis13,wojno17}}.
 As later we account for the incompleteness of the selected sample, we considered only a special region on the sky where the 
 selection function of the TGAS catalogue is defined \citep{bovy17a}. %Because of this we lose $\sim$70\% of the stars lying above the Galactic plane. 
 %Together with the other cuts listed below this considerably reduces the number of stars above the midplane. 
 We further selected only stars below the Galactic plane, $\tilde{z}<0$, as after all the cuts listed in this section 
 the fraction of stars left above the midplane is only 6.6\% of the total final sample, which can be safely neglected in order to 
 reduce the modelled volume and thus speed up the calculations.
 We also restricted ourselves to the local cylinder by applying $\tilde{d}\cos{b}<300$ pc and $|\tilde{z}|<1$ kpc.
 Hereafter we drop modulus in our notation of height and recall that our model is plane-symmetric and we work in the region below the midplane.
 The number of stars left is 51,234. 
 
 \item \label{parcut}\textit{Parallax cut.} About $1.3\%$ of the stars in the TGAS$\times$RAVE cross-match have negative parallaxes 
 because of the  impact of large observational errors on astrometric solutions for faint and/or distant stars.
 As a straightforward conversion $1/\varpi$ is not applicable in this case, we further included only stars with positive parallaxes, $\varpi>0$. 
 Next we selected stars with a relative parallax error smaller than 30\%,
 $\sigma_{\varpi}/{\varpi}<0.3$. After this cut, 50,491 stars remained.
 
 \item \label{photocut}\textit{Photometric cut.} We selected stars with known (B-V) APASS colour  
 that belonged to the range of visual magnitudes and colours prescribed by the observation windows of the RAVE and Gaia surveys: 
 \mbox{$7<I_{\mathrm{DENIS}}/\mathrm{mag}<13$}, \mbox{$0<(J-K_s)_{\mathrm{2MASS}}/\mathrm{mag}<1$} and \mbox{$0<J_{\mathrm{2MASS}}/\mathrm{mag}<14$}. 
 The remaining sample contains 45,478 stars. 
 
 \item \label{qualcut}\textit{Quality cut.} We set a lower limit to the signal-to noise ratio (S/N) of the RAVE spectra, $S/N>30$. 
 We also selected only stars  with $algo\_conv \neq 1$, which corresponds to the robust stellar parameters derived from the RAVE spectra. 
 Additionally, we ensured that only stars with 
 reliable values of TGAS astrometric parameters entered our sample. For this we applied the cuts to astrometric excess noise, 
 $\epsilon_i<1$, and its significance, $D[\epsilon_i]>2$. At this stage, 34,501 stars were left.
   
 \item \label{chemcut}\textit{Abundance cut.} To select the stars that can be identified as thin-disc members in chemical abundance plane,
 we first applied the cut $4000<T_{\mathrm{eff}}/K<7000$ as only for this range of effective temperatures the chemical abundances were determined in RAVE 
 \mbox{\citep{kunder17}}. Then the cuts $\mathrm{[Fe/H]}>-0.6$, $\mathrm{[Mg/Fe]}<0.2$ were added. As discussed in \mbox{\citet{wojno16}}, this is a reasonable criterion for the 
 separation of the thin and thick discs in the RAVE data.
\end{enumerate}

After applying all the selection criteria, we had a sample of 19,746 stars (Fig. \ref{data_sample}, blue). The numbers of stars given for each stage 
of the data cleaning should not be interpreted straightforwardly in terms of strictness of the applied criteria.
Because of the strong correlations between some of the criteria (e.g. cuts \ref{geomcut} and \ref{parcut}), 
these values are very sensitive to the order of applying the cuts. 
Alternatively, all the criteria might be sorted according to the origin of the 
quantities and the fractions of stars might be estimated that were removed from the initial sample due to the problems with TGAS or RAVE or even entries from TGAS and RAVE together. 
The corresponding fractions removed from the sample after a simple geometric pre-selection (300-pc radius, $-1000 \, \mathrm{pc}<\tilde{z}<0 \, 
\mathrm{pc}$) were 38\%, 14.5\%, and 22\%, respectively.

\subsection{Data incompleteness}\label{data_incompl}

By reducing the TGAS$\times$RAVE cross-match with criteria \ref{geomcut}-\ref{chemcut}, we selected 
a clean thin-disc sample in the local solar cylinder with high quality of the measured quantities. 
This final sample is not representative for the direct study of the 
vertical disc structure, however: the TGAS and RAVE catalogues are both incomplete, and the applied cuts exacerbate 
this incompleteness. To account for this, we thoroughly examined the sources of possible biases in our sample 
and constructed its selection function for further practical use. 

The final thin-disc sample clearly contains imprints of the selection functions from both of its parent catalogues. The selection function of the RAVE DR5 
depends in general on $I$-band photometry, $J-K_s$ colour, and position on the sky, $S_{\mathrm{RAVE}}(\alpha,\delta,I,J-K_s)$. 
However, the dependence on colour is relevant only for the low latitudes, $5^\circ<|b|<25^\circ$ \mbox{\citep{kordopatis13,wojno17}}
and is not of primary importance. The selection function of RAVE DR5, which was derived 
recently in \mbox{\citet{wojno17}} and is used in this paper, ignores the colour dependence, $S_{\mathrm{RAVE}}(\alpha,\delta,I)$. 
The completeness factor of the RAVE data varies strongly with the line of sight and visual $I$-band magnitude. 
The TGAS selection function was investigated in \mbox{\citet{bovy17a}} 
and was found to be quite homogeneous over a large part of the sky. 
It is defined in terms of 2MASS magnitudes and colours, \mbox{$S_{\mathrm{TGAS}}(\alpha,\delta,J-K_s,J)$}.
Because of the properties of the Gaia scanning law, the number of observations per star in Gaia DR1 is lowest near the 
ecliptic, that is, the measured quantities in the near-ecliptic regions are characterised by high uncertainties and are expected to be most biased by systematic errors. 
For this reason, near-ecliptic `bad' regions were excluded from the analysis in \mbox{\citet{bovy17a}}, such that 
the TGAS selection function is not defined for the whole sky. We considered only the region where the TGAS selection function is known (cut \ref{geomcut}).  

\begin{figure}
\centerline{\resizebox{\hsize}{!}{\includegraphics{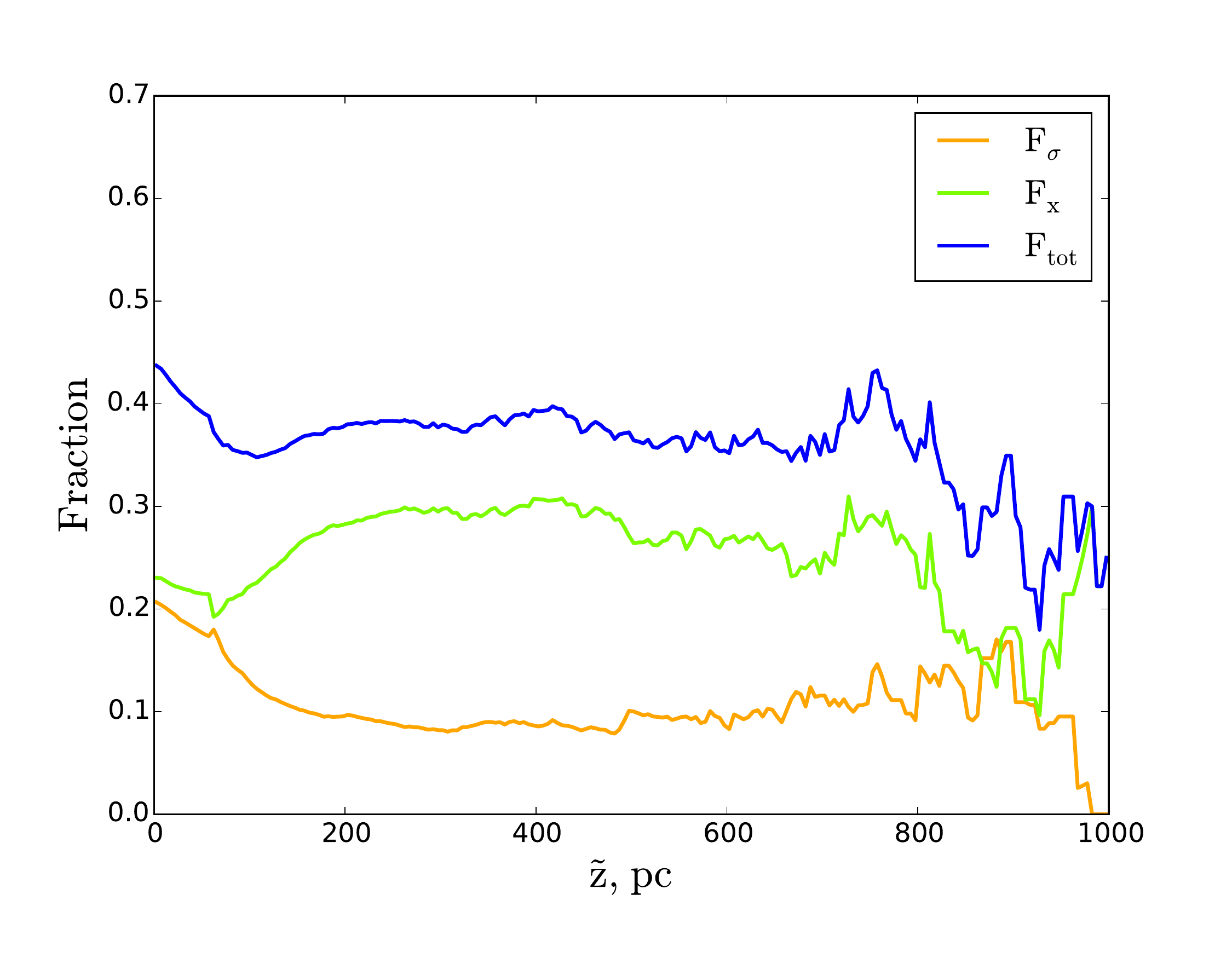}}}
\centerline{\resizebox{\hsize}{!}{\includegraphics{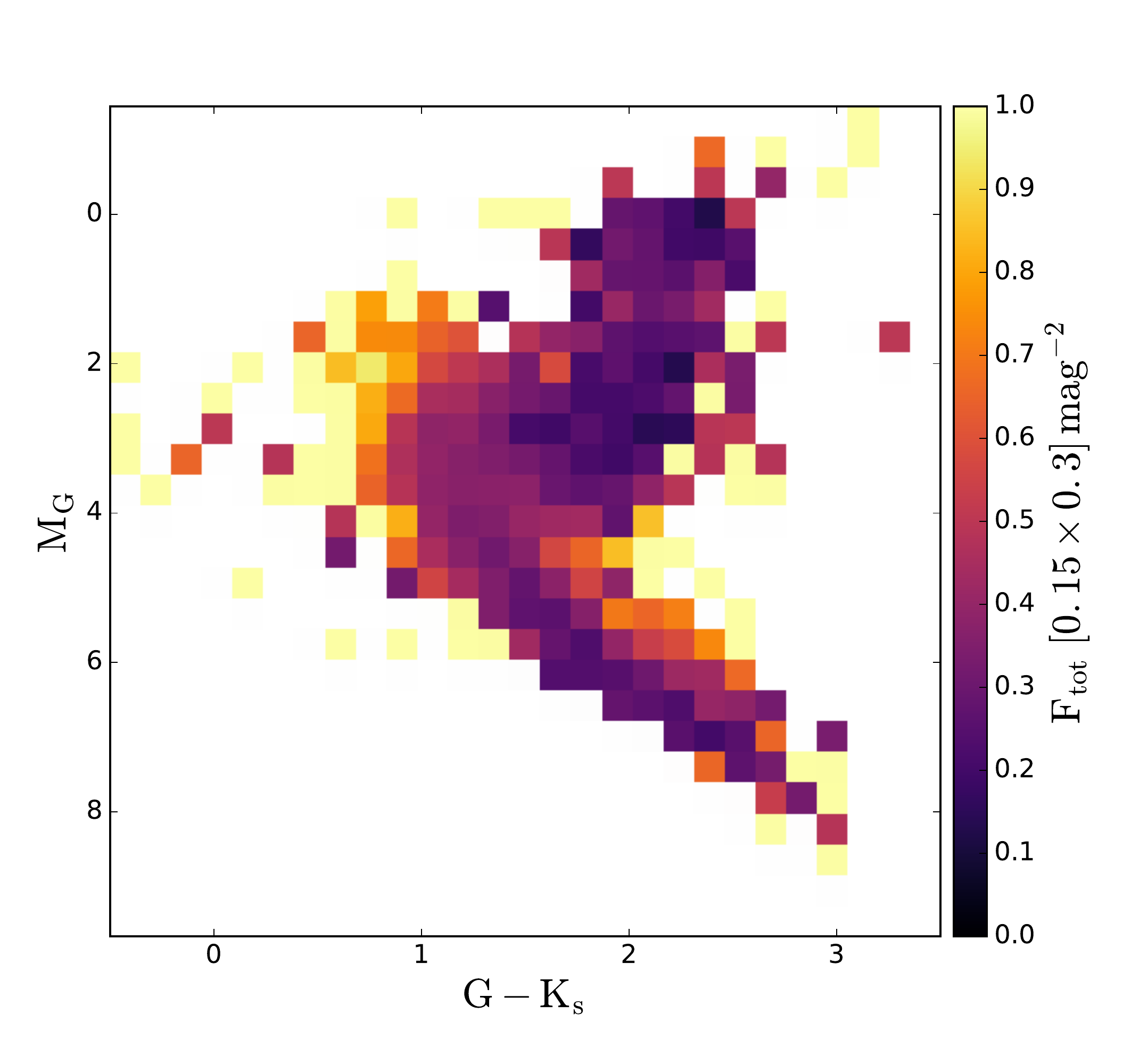}}}
\caption{\textit{Top}. Fraction of stars missing in the solar cylinder of 300 pc radius and 1 kpc height below the midplane. The orange line corresponds to the contribution 
from the low-quality stars, and the green curve gives an upper limit on the fraction of stars that did not enter the final sample 
because of the missing chemical abundances. The total fraction of missing stars is shown in blue.
\textit{Bottom.} Total fraction of the missing stars as a function of absolute G magnitude and $G-K_s$ colour. 
}
\label{missing_stars}
\end{figure}

Our selection criteria listed in Section \ref{cuts} also add to the total incompleteness of the final sample. 
The first cut specifies the geometry of the modelled volume and is of no interest here.  
The parallax cut was modelled as described in Section \ref{model_parallax_cut}. 
We also lost stars when applying the high-quality criteria (cut \ref{qualcut}) together with the selection of stars
with available APASS colour (part of cut \ref{photocut}). Finally, not all of the stars in the volume have 
measurements of Fe and Mg, even within the safe effective temperature range given. As a result, by applying cut \ref{chemcut}, we also 
removed a part of the upper main sequence (UMS) and lost thin-disc stars.
To quantify the fraction of the thin-disc stars missing in our final sample, we constructed two additional data sets. 
The first contains stars with known chemical abundances that are classified as thin-disc populations
but have low-quality or incomplete records (if any condition from cut \ref{qualcut} is not fulfilled or the APASS (B-V) colour is missing). 
The second data set contains stars that are unclassified as a result of missing chemical abundances, regardless of the quality of their spectra and astrometric solution or
available photometry. 
At this point, we can derive a reliable estimate of the missing star fraction. To determine the total number of the thin-disc stars in some vertical bin 
that also pass our parallax cut, we summed all the three samples in the following manner:
\begin{equation}
\label{td_fr1}
N_{tot} = N_{f} + N_{\sigma} + N_{x} \cdot \frac{ \sum_{\tilde{z}_{min}}^{\tilde{z}_{max}} \rho_d}{\sum_{\tilde{z}_{min}}^{\tilde{z}_{max}} (\rho_d+\rho_t)},
\end{equation}
where $N_f$, $N_{\sigma}$, and $N_x$ are the number of stars in the final, low-quality, and unclassified samples. The densities $\rho_d$ and $\rho_t$ are the local 
vertical density profiles of the thin and thick discs as inferred in \citetalias{just10} and \citetalias{just11}, respectively. 
The two limits $\tilde{z}_{min}$ and $\tilde{z}_{max}$ are the boundaries of the corresponding vertical bin. In order to simplify the expression, 
we dropped the dependence on $\tilde{z}$ of all quantities in this equation.
With this estimate we calculated the total fraction of the missing stars $F_{tot}$ as a function of height:
\begin{eqnarray}
\label{td_fr2}
F_\sigma &=& N_{\sigma}/N_{tot}; \ \ \  F_x = N_x/N_{tot} \\ \nonumber 
F_{tot} &=& (N_{\sigma}+N_x)/N_{tot}
\end{eqnarray}
Here $F_\sigma$ and $F_x$ are the fractions of stars that are missing in our final clean thin-disc sample because of the high noise and the absence of chemical abundances, respectively.
The fractions given by Eq. \ref{td_fr2} are shown in the top panel of Fig. \ref{missing_stars}. All three curves are smoothed with a window of 10 pc width. 
After summation over all vertical bins, we find that $\sim$37\% of the total expected number $N_{tot}$ are missing in our final sample. 
%Note, that the effect of missing stars is most important close to the Galactic plane, thus we expect this region to be most difficult to model reliably.
 
We also examined the fraction of missing stars in the Hess diagram. The bottom panel of Fig. \ref{missing_stars} 
with the $F_{tot}$ ratio calculated in colour-magnitude bins for the whole cylinder shows that the unclassified stars mostly influence the UMS; 
the LMS and red giant branch (RGB) regions are affected to a somewhat smaller extent. 
As it is not fully clear which of the unclassified stars really belong to the thin disc (the ratio of the thin- and thick-disc densities presented in Eq. \ref{td_fr1}
was used only to estimate the number of stars, it gives us no clue which population an individual star belongs to), 
we did not attempt to use the derived weights of the Hess diagram as an additional selection function during the modelling procedure. We rather kept it as a 
key for understanding the nature of the discrepancies between the data and model when they arose (see Section \ref{discus}).

Now, treating all the described selection effects as independent, we defined a completeness factor for the TGAS$\times$RAVE thin-disc sample as follows: 
\begin{eqnarray}
S = S_{Q}(\tilde{z}) \times S_{\varpi}(\tilde{z}) &\times& S_{\mathrm{RAVE}}(\alpha,\delta,I) \times \nonumber\\
\label{Sfunc}  &\times& S_{\mathrm{TGAS}}(\alpha,\delta,J,J-K_s). 
\end{eqnarray}
Here $S_{\varpi}(\tilde{z})$ corresponds to the selection effect arising from the parallax cut (practical realisation described in Section \ref{model_parallax_cut}).
$S_{Q}(\tilde{z}) = 1 - F_{tot}(\tilde{z})$ is the additional completeness factor related to the missing stars problem.
All the factors of the selection function except for $S_{Q}$ were applied during the creation of the stellar populations and calculation of such quantities of interest 
as Hess diagrams or velocity distribution functions, whereas $S_{Q}$ was taken into account at the post-processing of the results (Section \ref{model}).

\section{Model and simulations}\label{model}

\subsection{Basics of the JJ model}\label{basics}

Our self-consistent chemodynamical model of the thin disc has been presented and discussed in a series of papers. 
We refer to our original \citetalias{just10} for the model details and also for \citetalias{just11} and \citetalias{rybizki15} 
for its later improvement.  
Here we briefly review the model basics that are relevant for the further discussions.

The thin disc is constructed from a set of 480 isothermal monoage subpopulations 
with ages in the range of $\tau=0...12$ Gyr with an equal step of $\Delta \tau=25$ Myr. 
The dynamical heating of the subpopulations is given by the AVR function, which 
describes an increase in vertical velocity dispersion with stellar age as a result of the response to small perturbations in the gravitational potential. 
We model the AVR as a power law with the vertical velocity dispersion starting at $\sim$5 km s$^{-1}$ for the newly born stars and increasing up to 25 km s$^{-1}$ 
for the oldest subpopulation. 
Another input function of interest is the SFR, which is implemented as a two-parametric analytic function. When combined with the AVR and IMF, it is a 
sensitive tool to predict star counts and age distributions. 
A local chemical enrichment in terms of a monotonously increasing AMR is added to the model in order to reproduce the observed metallicity distributions. 
In addition to the thin disc, other Galactic components are included in the total mass budget. The gas is modelled with the scaled thin-disc AVR, 
its scale height was optimised to $h_g=100$ pc in order to reproduce the observed surface density of the gas. 
The thick disc is included as a single-birth population with an age of 12 Gyr. The spherical isothermal DM halo is added in a simple thin-disc approximation.
The thin-disc subpopulations are assumed to be in dynamic equilibrium in the total gravitational potential generated by the stellar, gaseous, and DM components of the Galaxy. 
By solving Poisson's equation iteratively, we obtain a self-consistent vertical potential and density laws of the thin-disc subpopulations.
Originally, the AVR parameters were calibrated against the kinematics of MS stars from the 
\textit{Hipparcos} catalogue combined at the faint end with the Fourth Catalogue of Nearby Stars (CNS4, \mbox{\citealp{jahreiss97}}). 
To constrain the AMR, the local metallicity distribution from the Copenhagen F and G star sample (GCS1, \mbox{\citealp{nordstrom04}}) was used \citepalias{just10}. The 
thin-disc vertical profile is essentially non-exponential close to the plane, characterised by half-thickness $h_d=400$ pc. 
However, kinematic data alone do not help to distinguish SFR and IMF. Further comparison of the model to the Sloan Digital Sky Survey (SDSS) star counts 
towards the north Galactic pole allowed us to distinguish the SFR and IMF \citepalias{just11} and constrain both of them \citepalias{just11,rybizki15}. 
Our best SFR has a peak at $\tau \approx 10$ Gyr and is characterised by the mean and present-day star formation rate of 
3.75 $\mathrm{M_\odot \ pc^{-2} \ Gyr^{-1}}$ and 1.4 $\mathrm{M_\odot \ pc^{-2} \ Gyr^{-1}}$, correspondingly.
At this stage, the thick-disc parameters were also improved. The local surface density was set to 5.4 $\mathrm{M_\odot \ pc^{-2}}$ (18\% of the thin-disc surface density), 
the scale height was fixed to $h_t = 800$ pc, and the vertical velocity dispersion was $\sigma_t=45.4$ km s$^{-1}$.

We note that although numerous recent studies indicate that the thick disc might be more complex than we adopt here 
(e.g. \mbox{\citealp{bovy12td1,bovy12td2,haywood13}}), 
the topic is still under debate. It is generally recognised that the thick disc is an old, alpha-rich, and metal-poor population 
formed on a short timescale. 
The span of the suggested timescale varies significantly in different studies 
(4-5 Gyr in \mbox{\citealp{haywood13}}, but less than 2 Gyr with as low as 0.1 Gyr in the two-infall chemical models in 
\mbox{\citealp{grisoni17,grisoni18}}). 
Some authors \mbox{\citep{haywood13,hayden17}} reported an increase in vertical velocity dispersion with the age of thick-disc populations. 
On the other hand, 
within the framework of the Jeans analysis developed in \mbox{\citet{sysoliatina18}}, 
the alpha-rich metal-poor thick disc behaves as a well-mixed kinematically homogeneous population. 
As the local data alone are not enough to distinguish the thin and thick discs in a robust way, 
we did not attempt to re-define them in this work. For a quick test against Gaia DR2 (see end of Section \ref{discus}), 
we modelled the thick disc with a metallicity spread. Further possible improvements are postponed to the stage when the model is fully extended,
which will also include the detailed chemical evolution. This will allow us to study different Galactic populations in the abundance plane.

Thus, at the moment we arrived at a monotonous AVR and AMR, a decreasing SFR, and a three-slope broken power-law IMF 
for the thin disc and a single-age, scale height, and vertical velocity dispersion thick disc.
Following the convention established in \citetalias{just10}-\citetalias{rybizki15}, 
we address this best set of parameters as the fiducial model A and use it for the modelling throughout this paper. 

\subsection{Modelling procedure}\label{modelling}

\begin{figure*}[ht!]
\centerline{\resizebox{1\hsize}{!}{\includegraphics{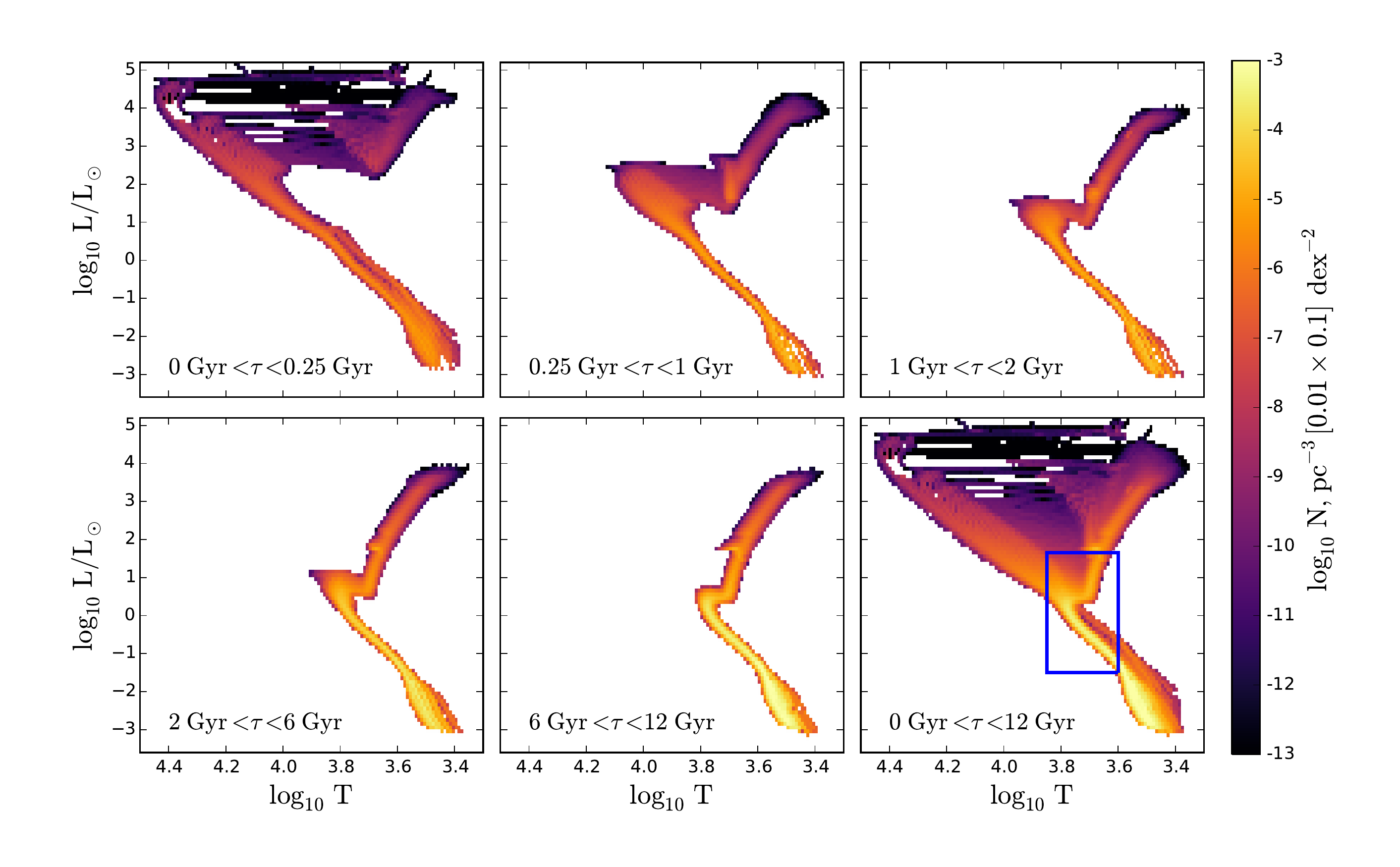}}}
\caption{Theoretical HR diagrams for the thin disc as predicted by the local JJ model with the use of MIST isochrones.
Five age bins illustrate the contributions from the different thin-disc populations to the total HR diagram (lower right). The blue box 
roughly corresponds to the region studied in this work.}
\label{N_in_volume_plot}
\end{figure*}

Before going into the details, we sketch a general overview of the modelling procedure summarising all steps discussed below in Sections \ref{table}-\ref{verted}.

We start with discretisation of the age-metallicity space and assigning an isochrone to each age-metallicity pair (Section \ref{table}).
After this, we create a cylindrical grid (Section \ref{grid}) and populate each of its space volumes with the predicted types of stars 
(characterised by age, metallicity, and mass from the isochrone) and  
take into account the parallax cut (Section \ref{model_parallax_cut}), reddening, the TGAS$\times$RAVE selection function (Section \ref{incompl}), 
and the corresponding abundance and photometric cuts. Then we account for the vertical effect of the distance error (Section \ref{verted}) 
and investigate the properties of the mock sample as a function of distance from the midplane (Section \ref{results}). 

\subsection{Creation of the stellar assemblies}\label{table}

We started our modelling with thin-disc mono-age subpopulations. Each age corresponds to a unique value of metallicity as prescribed by the AMR law. 
We added a Gaussian scatter in metallicity $\sigma_{\mathrm{[Fe/H]}}=0.15$ dex. 
Each age $\tau_i$ was then associated with seven metallicities representing a Gaussian with a mean AMR$(\tau)$ and a dispersion $\sigma_{\mathrm{[Fe/H]}}$.
Each metallicity subpopulation is given an appropriate weight $w_k$, such that $\sum_{k=1}^7 w_k = 1$. 
In \citetalias{just10} a similar scatter in metallicity was modelled to reproduce the observed metallicity 
distributions of Geneva-Copenhagen F and G stars, which can be interpreted either in terms of observational errors or in terms of real physical scatter in metallicity 
of mono-age populations. The added dispersion $\sigma_{\mathrm{[Fe/H]}}$ here corresponds to a physical scatter in metallicity as the age-metallicity 
grid was then used to select isochrones. This complexity roughly accounts for the effect of radial migration, which is not explicitly introduced in our model.

At this point, it is useful to define a new notation in order to avoid confusion in the future. From here on, we refer with `subpopulation'  
to the thin-disc mono-age isothermal subpopulations. For the additional splitting in metallicity-mass parameter space we introduce the term `stellar assembly'.
Thus, the described splitting of the subpopulations in metallicity results in $480 \cdot 7 = 3360$ stellar assemblies.

The next step is to include the full stellar evolution. 
We used the Modules and Experiments in Stellar Astrophysics (MESA, \citealp{paxton11,paxton13,paxton15}) 
Isochrones and Stellar Tracks (MIST, \citealp{dotter16,choi16})\footnote{\url{http://waps.cfa.harvard.edu/MIST}}. 
We took a set of 67 isochrone tables that safely cover the whole range of the modelled metallicities, $\mathrm{[Fe/H]}=-0.6...0.46$.  
The ages of isochrones are given in logarithmic range $\log{(\tau/\mathrm{yr})}= 7...10.08$ with a step of 0.02. 
Each of the 3360 stellar assemblies constructed earlier 
was associated with the isochrone that had the closest age and metallicity to the modelled ones.
The MIST isochrones cover a range of initial masses $0.1...300 \ M_{\odot}$ from which we selected the range $0.1...100 \ M_{\odot}$
as prescribed by the model IMF. 
The resulting number of the stellar assemblies characterised now by the same ages, metallicities, and initial stellar masses is $\sim 5 \cdot 10^6$.
For each stellar assembly $j$, a surface number density $N^\Sigma_j$ was calculated by weighting the SFR with the IMF and accounting for the weights $w_k$. 
Isochrones also provide us with the present-day stellar masses, stellar parameters $\log{L}$, $\log{T_{\mathrm{eff}}}$, $\log{g}$, and absolute magnitudes 
in the standard $UBVRI$ system, 2MASS $JHK_s$, and Gaia $G$ band. 
% precise number for PARSEC: 1550788. 
At this step we have a table of the mock stellar assemblies with their known surface densities in the solar neighbourhood. 
The remaining modelling procedure may be briefly described as checking which of these types of stars were actually observed in the TGAS$\times$RAVE cross-match 
and then were selected by us for the thin-disc sample as well as for further distributing them in the local cylinder. 

The volume number density is calculated by:
\begin{equation}
N^V_{j}(z) = \frac{N^{\Sigma}_j}{2 h(\tau_j)}\exp{\big\{- \Phi(z)/\sigma_W^2(\tau_j)\big\}},
\label{N_in_volume}
\end{equation}
where $N^{\Sigma}_j$ is the surface number density of a given stellar assembly, $h(\tau_j)$ and $\sigma_W(\tau_j)$ 
are the half-thickness and the vertical velocity dispersion defined by the AMR for the corresponding age $\tau_j$, respectively, and $\Phi(z)$ 
is the total vertical potential at a height $z$, as predicted by the fiducial model A. 
As an example outcome of the local model, we calculated the volume densities $N_{V}$ at $z=0$ kpc for all modelled stellar assemblies 
and constructed a Hertzsprung-Russel (HR) diagram for the thin disc in the solar neighbourhood (Fig. \ref{N_in_volume_plot}). 

Before proceeding to the next steps of the modelling, we pre-selected stellar assemblies with the effective temperatures 
belonging to the RAVE range (cut \ref{chemcut}). 
An additional pre-selection was based on the expectations for the range of absolute G magnitudes 
in the sample, which we set to $-2...10$ mag (see bottom panel of Fig.\ref{missing_stars}). 
The reduced subset of the stellar assemblies used further during the simulations falls into the
blue frame on the bottom right plot in Fig. \ref{N_in_volume_plot}.

\subsection{Sample geometry}\label{grid}

\begin{figure}[t]
\centerline{\resizebox{1\hsize}{!}{\includegraphics{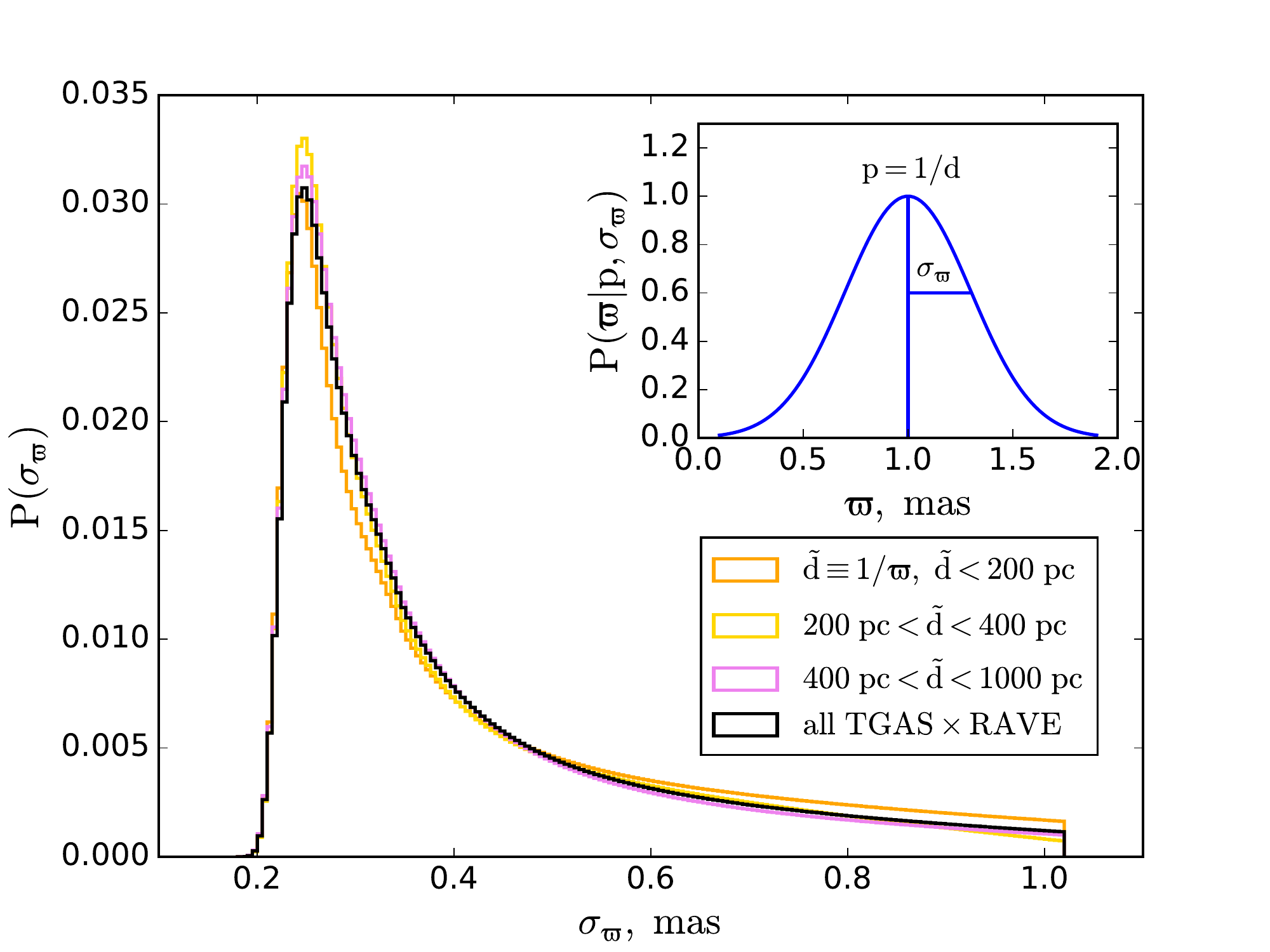}}}
\caption{Normalised parallax error PDF as taken from the full TGAS$\times$RAVE cross-match (black).
The orange, yellow, and violet histograms correspond to the PDFs of the subsamples with different distance ranges. 
The similarity of their shapes demonstrates an independence of the parallax error from the parallax itself (i.e. the distance).
In the inset an example normalised Gaussian PDF for the observed parallax is plotted. 
The true parallax and its observational error are set to 1 and 0.3 mas, respectively.}
\label{epar_plot}
\end{figure}

To model the geometry of the thin-disc sample, we created a 3D grid in cylindrical coordinates $(r,\theta,z)$. 
To adequately account for the parallax cut (see below), we
set the initial radius of the cylinder to $r_{in}=500$ pc. This is larger than we used for the data (cut \ref{geomcut} in Section \ref{cuts}). 
The modelled cylinder extends vertically up to $z_{max}=1$ kpc below the midplane. The region above the Galactic plane was not included. 
We binned this cylindrical volume (1) in height $z$ with a step of $\Delta z = 5$ pc, (2) 
in angle $\theta$, which was measured from the cylinder axis, with a step of $\Delta \theta = 3^\circ$, and (3) in radial direction $r$
by binning the space in 20 intervals in logarithmic scale. 
This gives $4.8\cdot10^5$ space volumes, but only those of them that correspond to the directions of observations 
were used in the actual modelling.

\subsection{Parallax cut}\label{model_parallax_cut}

The distances $r$ and $z$ defining the model grid are not directly comparable to the distance estimates derived for the data sample 
through the inversion of observed parallaxes. 
As a work-around, we translated the model true distances into the space of observed parallaxes. 
This allowed us to include in the model the same parallax cut as was applied to the data, and also to account for the distance error effects.

We split the procedure into two independent steps. First, we applied Eq. \ref{parpdf} at fixed z and translated the model distances to the observed parallaxes. 
This allowed us to reduce the number of the stellar assemblies left for the modelling at a given height by applying a
parallax cut identical to the one we introduced in Section \ref{cuts}. Then we reduced the radius of the modelled cylinder to 300 pc 
using the distances calculated from the mock observed parallaxes. 
At this stage, we treated different z-slices as independent, that is to say, the modelled stellar assemblies 
remained at their initial height. The fact that a given star is observed at $\tilde{z}$ different from its true distance from the midplane z 
was taken into account at the second stage during a post-processing of the quantities calculated in the different z-slices. 
This strategy was chosen in order to simplify the modelling process and speed up the calculations, as it allowed us to include the effect of the distance error 
in vertical direction at the post-processing stage of the simulations. 
We now consider the first step with the parallax cut more closely; the second step is described in Section \ref{verted}. 

We started with the usual assumption that the observed parallax $\varpi$ is distributed normally
about the true parallax $p=1/d$ with a standard deviation $\sigma_{\varpi}$ depending on the stellar brightness, exposure time, and number of observations per star. 
This is a good approximation of the real shape of the TGAS parallax error distribution which is known to deviate slightly from normality only 
beyond $\sim2\sigma$ \mbox{\citep{lindegren16}}. Under this assumption, the normalised probability of the observed parallax $\varpi$ is given by the Gaussian PDF: 
\begin{eqnarray}
\label{parpdf} 
P(\varpi|\sigma_{\varpi},p) &=& \frac{1}{A} \exp{\Big[ - \frac{(\varpi - p)^2}{2 \sigma_{\varpi}^2}  \Big]} \\ \nonumber 
\mathrm{with} \ A &=& \sqrt{2 \pi} \sigma_{\varpi}. 
\end{eqnarray}
An example PDF of the observed parallax is shown in the inset of Fig. \ref{epar_plot}.  

At each fixed height z, we have 2400 volume spaces, as implied by the model grid. We assumed that the parallax error PDF as 
derived from the TGAS$\times$RAVE cross-match (Fig. \ref{epar_plot}, black histogram) does not depend on parallax or direction of observation 
(see coloured PDFs in Fig. \ref{epar_plot}). 
After this, we can easily assign parallax errors to our stellar assemblies when modelling the stellar content in each volume space. 
Then the observed parallaxes can be derived from the Gaussian distribution in accordance with Eq. \ref{parpdf}. We performed this  
separately for each radial bin in a given z-slice.
In summary, at each fixed z, we selected a pie of 20 space volumes corresponding to the same angle $\theta$ and different distances from the cylinder axis, and 
in each of them, we reduced the table of stellar assemblies in the following way: 
\begin{itemize}
\renewcommand{\labelitemi}{$\bullet$}
\item  A random true heliocentric distance $d$ within the given space volume was assigned to each stellar assembly and was then converted into the true parallax $p=1/d$.
\item  Parallax errors $\sigma_{\varpi}$ were drawn randomly from the parallax error PDF.
\item  The corresponding observed parallaxes $\varpi$ were derived as random values from the Gaussian PDF $P(\varpi|\sigma_{\varpi},p)$.
\item  The observed parallaxes $\varpi$ were converted into observed distances $\tilde{d}=1/\varpi$.
\item  Cuts identical to those used for the data sample were applied: $\varpi>0$, $\sigma_{\varpi}/\varpi<0.3$.
\item  Only stellar assemblies with $\tilde{d} \cos{b}<300$ pc were selected. This allowed us to account for the fact that the parallax errors 
cause some stars to lie outside the modelled cylinder, but may yet be observed within it (alternatively, stars that actually belong to the cylinder of 300 pc radius may 
appear at larger distance and be excluded from the sample).
\item  We also removed the stellar assemblies located at $|b|<20^\circ$ to reproduce the spatial geometry of the modelled sample.
\end{itemize}
The resulting set of tables of the stellar assemblies were then subjected to the TGAS$\times$RAVE incompleteness factor.

\subsection{TGAS$\times$RAVE selection function}\label{incompl}

After the parallax cut was applied, in each space volume at a given $z$ 
we calculated the visual magnitudes in the $I$ and $J$ bands as well as colour $(J-K_s)$ 
for all the stellar assemblies that were left. Apparent or absolute magnitudes in other photometric bands such as $B,V$, or $G$ were also added when necessary. 
Visual magnitudes were calculated with the true model distances $d$. 

The recent progress in mapping the dust content of the Milky Way allowed us to use a fully realistic 3D extinction model 
instead of the 2D map of \mbox{\citet{schlegel98}} \citepalias{just11} or the simple analytic extinction model from 
\mbox{\citet{vergely98}} \citepalias{rybizki15}. We implemented the 3D dust map from 
\mbox{\citet{bovy15}}\footnote{\url{https://github.com/jobovy/mwdust}},
which is a combination of 3D extinction models from \mbox{\citet{drimmel03}}, \mbox{\citet{marshall06}}, and \mbox{\citet{green15}}. 

The photometric cuts \mbox{$7<I/\mathrm{mag}<13$}, \mbox{$0<(J-K_s)/\mathrm{mag}<1$}, and \mbox{$0<J/\mathrm{mag}<14$} 
were then applied to the stellar assemblies, in full analogy with the corresponding data selection criterion (cut \ref{photocut}, Section \ref{cuts}). 
The surface number densities of the remaining stellar assemblies were weighted with the completeness factor $S_{\mathrm{RAVE}}\times S_{\mathrm{TGAS}}$. 
For this we binned the stellar assemblies in $I$ and $J$ magnitudes with the corresponding steps of $\Delta I=0.5$ mag
and $\Delta J=0.1$ mag as well as in colour $(J-K_s)$ with a $\Delta (J-K_s)=0.14$ mag step. 
After this, we calculated the resulting stellar number in a given grid cell 
by converting the surface stellar densities $N^{\Sigma}_j$ into $N^V_j$ in accordance with Eq. \ref{N_in_volume} 
and multiplying the volume density by the volume of the cell. With the known number of stars
as well as their multi-band photometry, ages, and metallicities we can predict the vertical density laws for different populations, 
calculate vertical kinematics
with the AVR, and study the modelled sample in the colour-magnitude plane. 

\subsection{Vertical distance error}\label{verted}

\begin{figure}[ht]
\centerline{\includegraphics[scale=0.4]{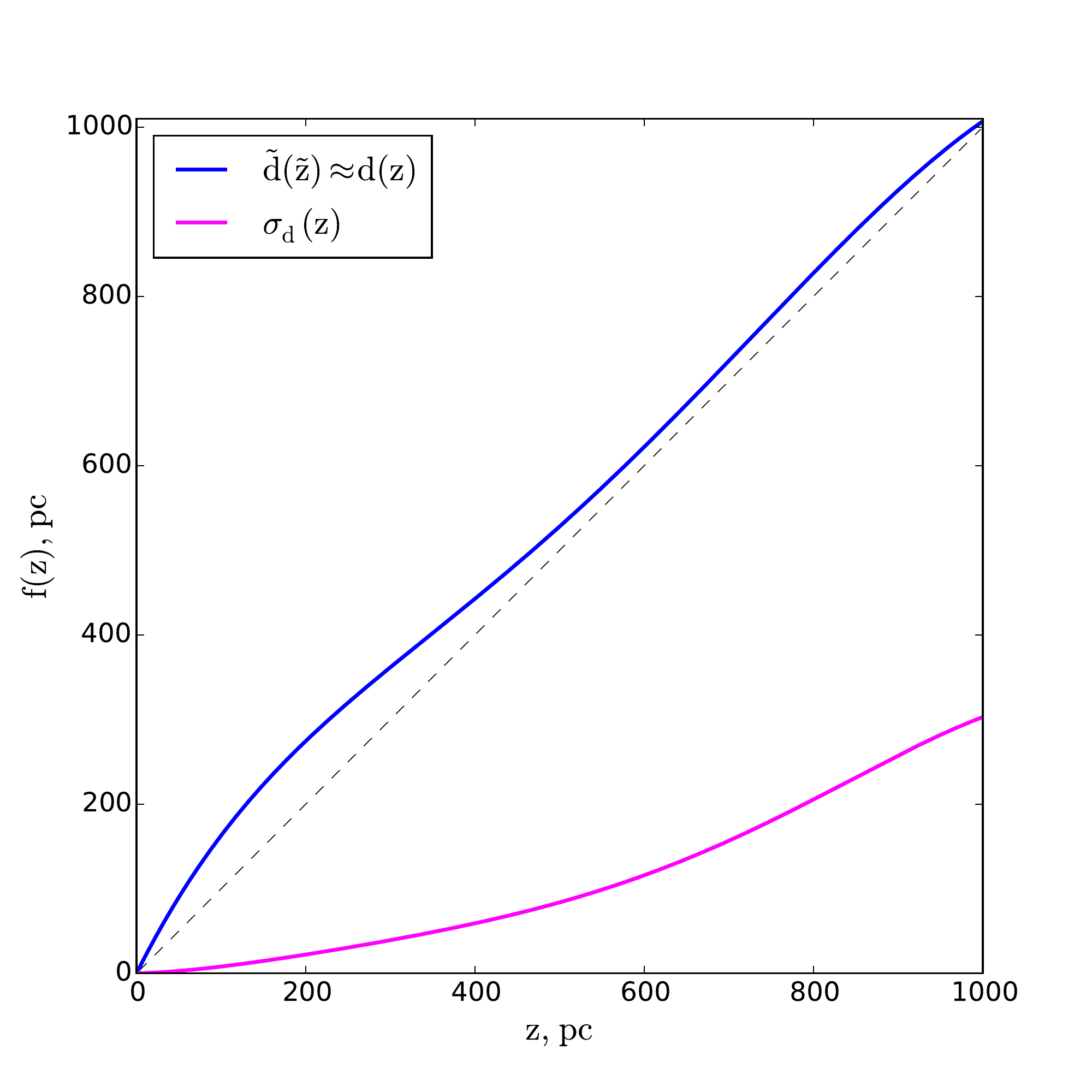}} 
%\centerline{\includegraphics[scale=0.4]{Vert_ed.pdf}} % referee mode
\centerline{\resizebox{1\hsize}{!}{\includegraphics{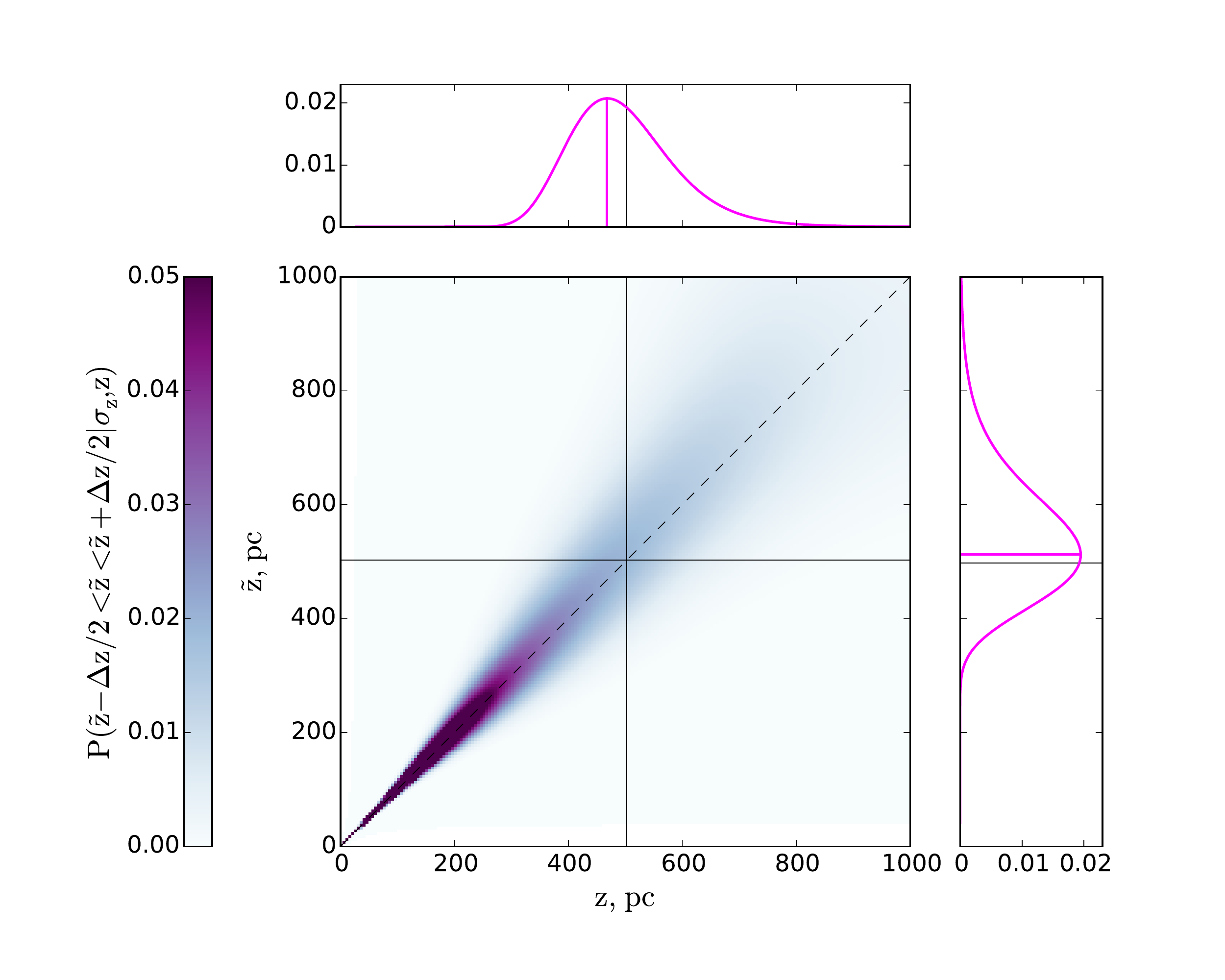}}}
\caption{\textit{Top.} Mean observed heliocentric distance $\tilde{d}$ vs. height $\tilde{z}$ as derived from the data (blue curve).
The thin-disc sample was binned in $\tilde{z}$, and at each height, the mean heliocentric distance was calculated. The observed dependence was
fitted with a polynom of the fifth order to obtain a smooth function. The distance error as a function of height was then derived with the 
derived mean distance (magenta curve). %\newline
\textit{Bottom.} Diagram showing the vertical effect of the distance error. 
The normalised probability for a true height $z$ to be associated with the observed distance from the 
midplane $\tilde{z}$ given the thin-disc sample geometry and the typical distance errors from the top panel is colour-coded. The horizontal and vertical projections
are example PDFs $P(\tilde{z}|\sigma_z,z)$ and $P(z|\sigma_z,\tilde{z})$ for the true and observed heights of 0.5 kpc. The skewness
of the two distributions results in two effects: (1) an individual star is more likely to be observed at a height larger than its true distance from the midplane, and
(2) the observations at some $\tilde{z}$ are more likely to be affected by smaller heights. The magenta lines in the projection subplots 
show distribution maxima.}
\label{verted_plot}
\end{figure}
At the stage of post-processing the results, when the quantities of interest were calculated for different $z$, we also accounted for the effects of the 
distance errors in the vertical direction. To do so, we introduced a probability $P(\tilde{z}|\sigma_z,z)$, 
which defines a likelihood for a quantity $Q$ calculated at its true $z$ to be observed at another height $\tilde{z}$ 
given a corresponding vertical error $\sigma_z$. In order to derive the $P(\tilde{z}|\sigma_z,z)$ expression,
we rewrote Eq. \ref{parpdf} in terms of heliocentric distance: 
\begin{eqnarray}
P(\tilde{d}|\sigma_d,d) &=& \frac{1}{B} \exp{\Big[ - \frac{d^4}{\sigma_d^2} (1/\tilde{d} - 1/d)^2 \Big]} \\ \nonumber 
\mathrm{with} \ B &=& \int_{0}^{\infty} P(\tilde{d}|\sigma_d,d) d \tilde{d},
\label{distpdf}
\end{eqnarray}
where $\tilde{d}$ and $d$ are the observed and true heliocentric distances and $\sigma_d$ is the distance error. 
At this point, we need to know how both distances as well as the distance error correspond to the true and observed heights, that is, the functions of interest 
are $\tilde{d}(\tilde{z})$, $d(z)$ and $\sigma_d(z)$. The first can be easily derived from the thin-disc sample itself (blue curve in the top panel of 
Fig. \ref{verted_plot}). Its behaviour is ruled by the sample geometry. Owing to the cut $|b|>20^\circ$, close to the midplane the radius of the cylinder 
grows linearly with $\tilde{z}$ until it reaches 300 pc, so that the mean distance $\tilde{d}$ increases almost linearly with 
height. At large $\tilde{z}$, the mean distance asymptotically approaches $\tilde{z}$. The function $d(z)$ can be calculated within the model 
framework, but for our accuracy, it is sufficient to use an approximation $\tilde{d}(\tilde{z}) \approx d(z)$. Then the distance error 
is estimated as $\sigma_d(z)=1/d^2(z) \cdot \sigma_{\varpi}$ with a constant $\sigma_{\varpi}=0.3$ mas, which is a typical parallax error of the TGAS
catalogue (magenta curve in the top panel of Fig. \ref{verted_plot}). 

\begin{figure}[ht!]
\centerline{\resizebox{\hsize}{!}{\includegraphics{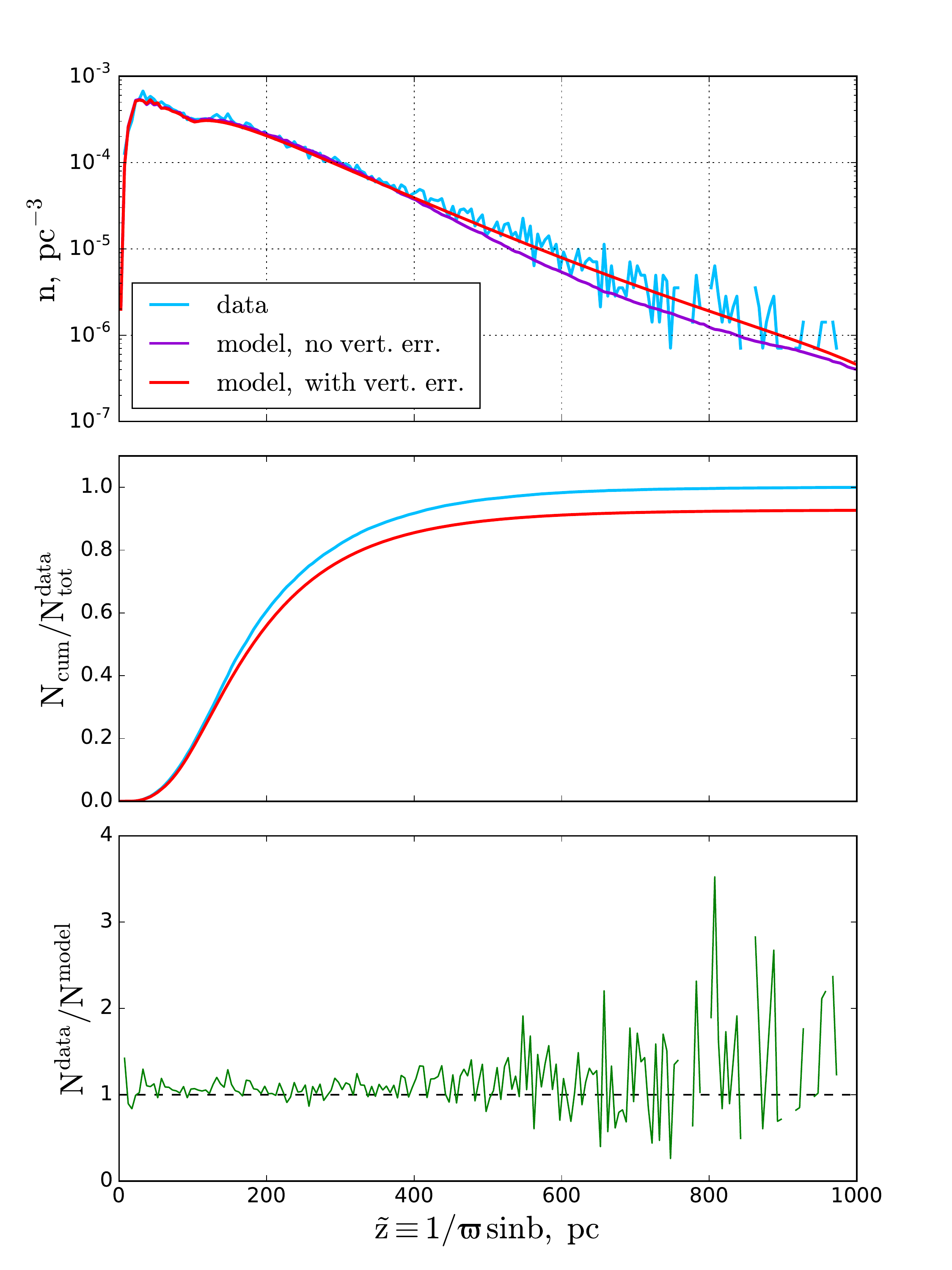}}}
%\centerline{\includegraphics[scale=0.4]{All_Nz.pdf}}
\caption{\textit{Top.} Volume number density of stars in the modelled cylinder. 
The blue line represents the data, and violet and red curves are the model 
predictions before and after the vertical error effect is taken into account.  
\textit{Middle.} Cumulative number of stars in the cylinder as a function of observed height $\tilde{z}$. 
The values are normalised to the total number of stars in the TGAS$\times$RAVE thin-disc sample.
\textit{Bottom.} Ratio of the observed to predicted number of stars in the vertical bins.}
\label{Nz_plot}
\end{figure}

The vertical effect of the distance error is illustrated in the bottom panel of Fig. \ref{verted_plot}. 
The colour-coding represents the normalised probability \mbox{$P(\tilde{z}-\Delta z/2<\tilde{z}<\tilde{z}+\Delta z/2|\sigma_z,z)$}.
The two projections of this diagram are of special interest. The vertical projection, calculated for a true height $z=500$ pc, 
gives a skewed PDF that describes a likelihood for a star located at $z$ to be observed at other heights $\tilde{z}$,
$P(\tilde{z}|\sigma_z,z)$. The horizontal projection 
shown for the same but observed height $\tilde{z}=500$ pc describes the impact from the different vertical bins on a given $\tilde{z}$, 
which is expressed with the PDF $P(z|\sigma_z,\tilde{z})$. 
While the maximum of the probability density $P(\tilde{z}|\sigma_z,z)$ belongs to a slightly larger height than the corresponding true $z$, the shift is opposite in case of  
$P(z|\sigma_z,\tilde{z})$: a star observed at a given $\tilde{z}$ most probably belongs in reality to a smaller height. It is also clear that 
the vertical effect of the distance error is negligible close to the midplane, where the distance errors are small, and has the largest impact 
on the location of the most distant stars. We also have to consider heights larger than $z_{max}=1$ kpc in order to allow the stars to not just leave 
the cylinder, but also to enter it from the larger heights (in full analogy with the radial direction, when we first consider a cylinder of a larger 
radius and then cut it to 300 pc after allowing stars to migrate horizontally due to the distance error effect). However, as the 
stellar density decreases approximately exponentially with increasing height, we expect the effect of not including $z>z_{max}$ to be very small 
except in the most distant parts of the cylinder, so we did not model it. 

\begin{figure}[ht!]
\centerline{\resizebox{\hsize}{!}{\includegraphics{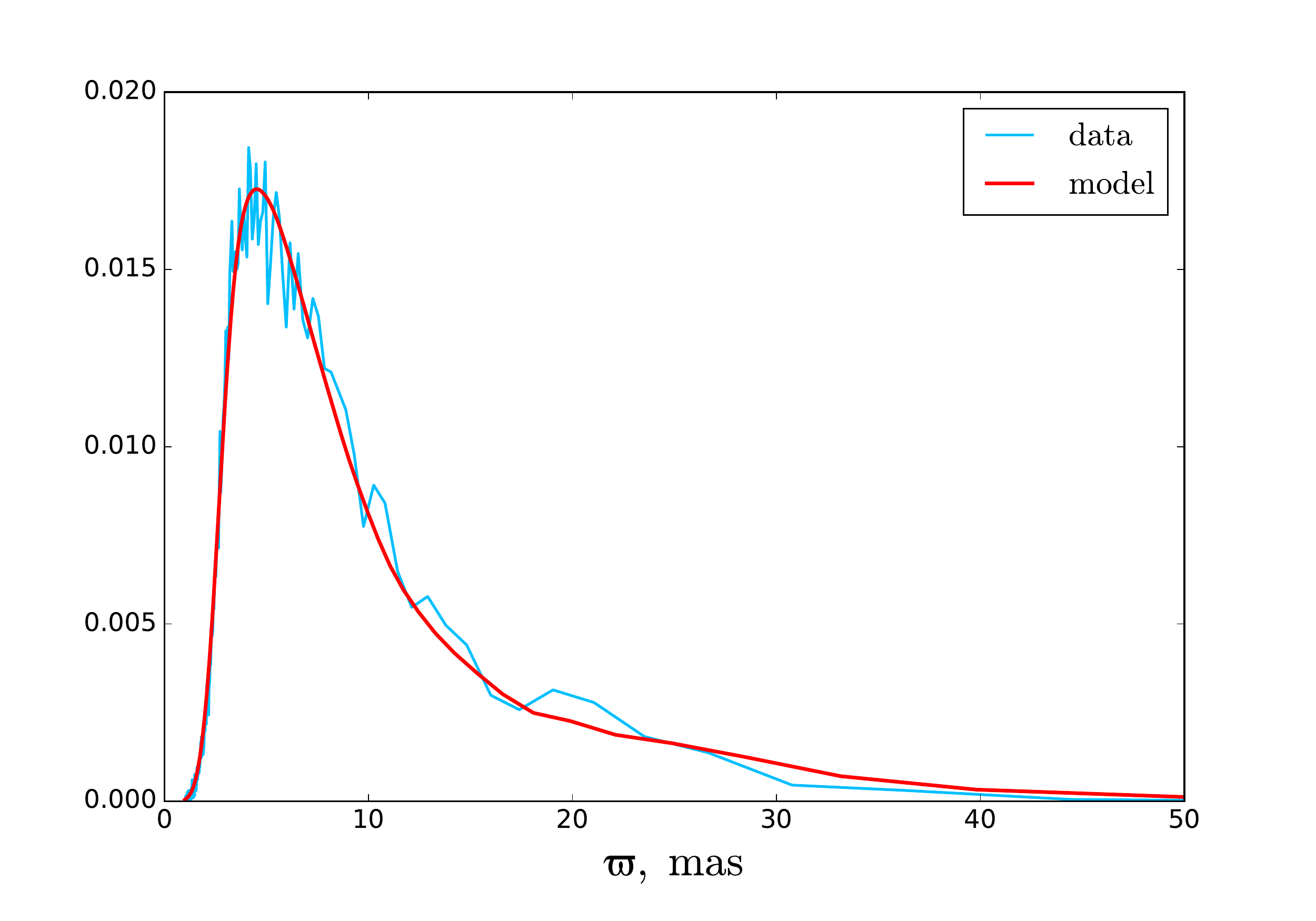}}}
%\centerline{\includegraphics[scale=0.37]{Par_dist.pdf}}
\caption{Normalised parallax distributions as derived from the TGAS$\times$RAVE thin-disc sample and as predicted by the JJ model.}
\label{Par_plot}
\end{figure}

Finally, we write
\begin{equation}
P(\tilde{z}|\sigma_z,z) = P(\tilde{d}(\tilde{z})|\sigma_d(z),d(z)). 
\label{Pz}
\end{equation}
With Eq. \ref{Pz} we calculated the quantities of interest as functions of observed height $\tilde{z}$:
\begin{equation}
Q_{\tilde{z}} = S_{Q} (\tilde{z}) \cdot \sum_{z=0}^{z_{max}} Q_{z} P(\tilde{z}|\sigma_z,z)
\label{Qz1}
\end{equation}
Here an additional and final correction was included in the form of the $S_{Q}(\tilde{z})$ factor discussed in Section \ref{data_incompl}.
Alternatively, when we wish to derive the model predictions in the horizontal slice of a thickness of $|\tilde{z}_2-\tilde{z}_1|$, we write 
\begin{equation}
Q_{\tilde{z_1} < \tilde{z} < \tilde{z_2}} = \sum_{\tilde{z}=\tilde{z}_1}^{\tilde{z}=\tilde{z}_2} Q_{\tilde{z}}.
\label{Qz2}
\end{equation}
Eq. \ref{Qz1} and \ref{Qz2} are used to account for the vertical effects of the distance error for the modelling of the stellar density laws and 
vertical kinematics. An alternative procedure for the Hess diagrams is presented in Section \ref{hess_results}. 

\section{Results}\label{results}
\subsection {Vertical density law}\label{Nz_results}

\begin{figure*}
\center{\includegraphics[scale=0.4]{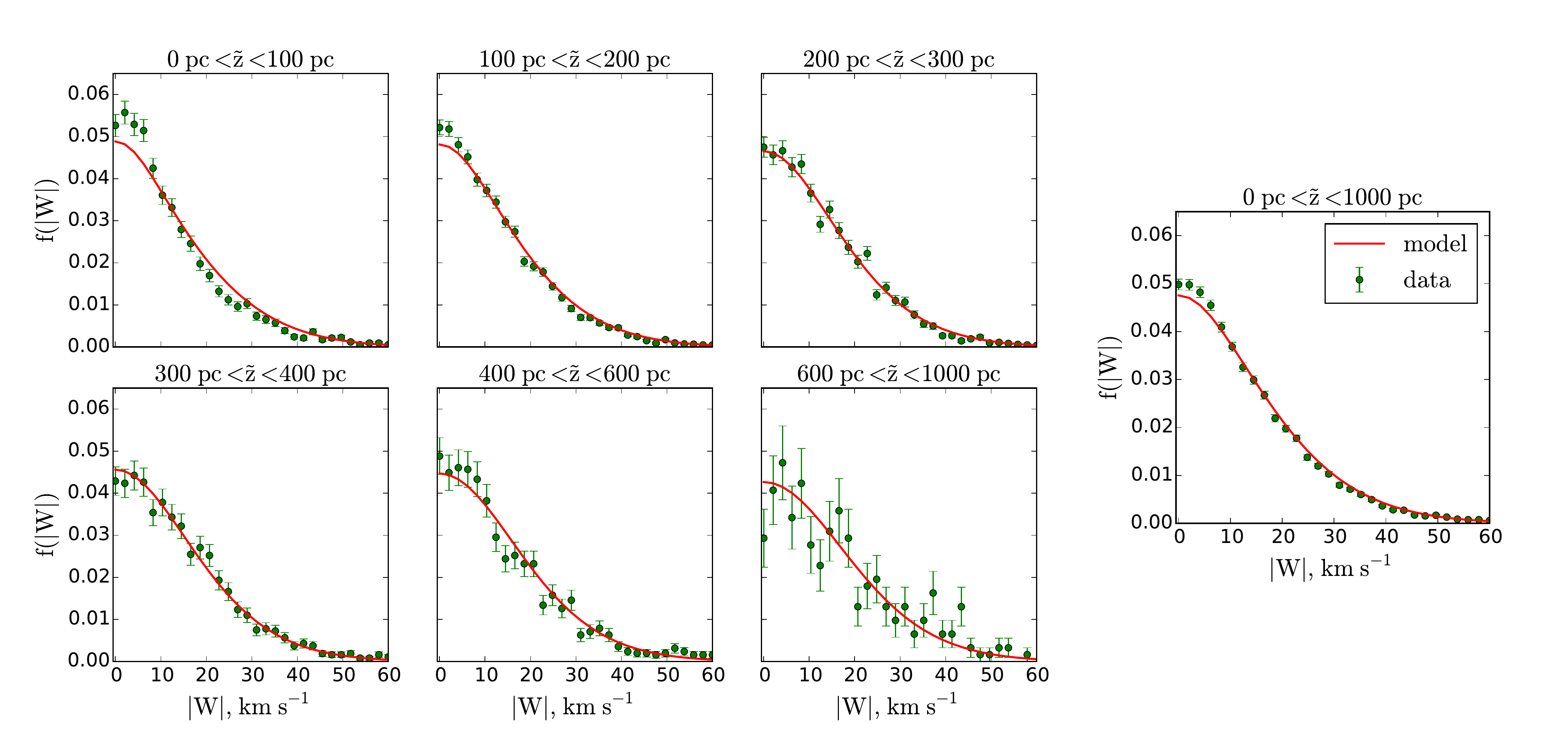}} % 0.4
\caption{Normalised velocity distribution functions $f(|W|)$ for the six horizontal slices (3$\times$2 plots on the left). 
The right graph shows $f(|W|)$ for the whole modelled cylinder.}
\label{fWz_plot}
\end{figure*}
\begin{figure*}
\center{\includegraphics[scale=0.4]{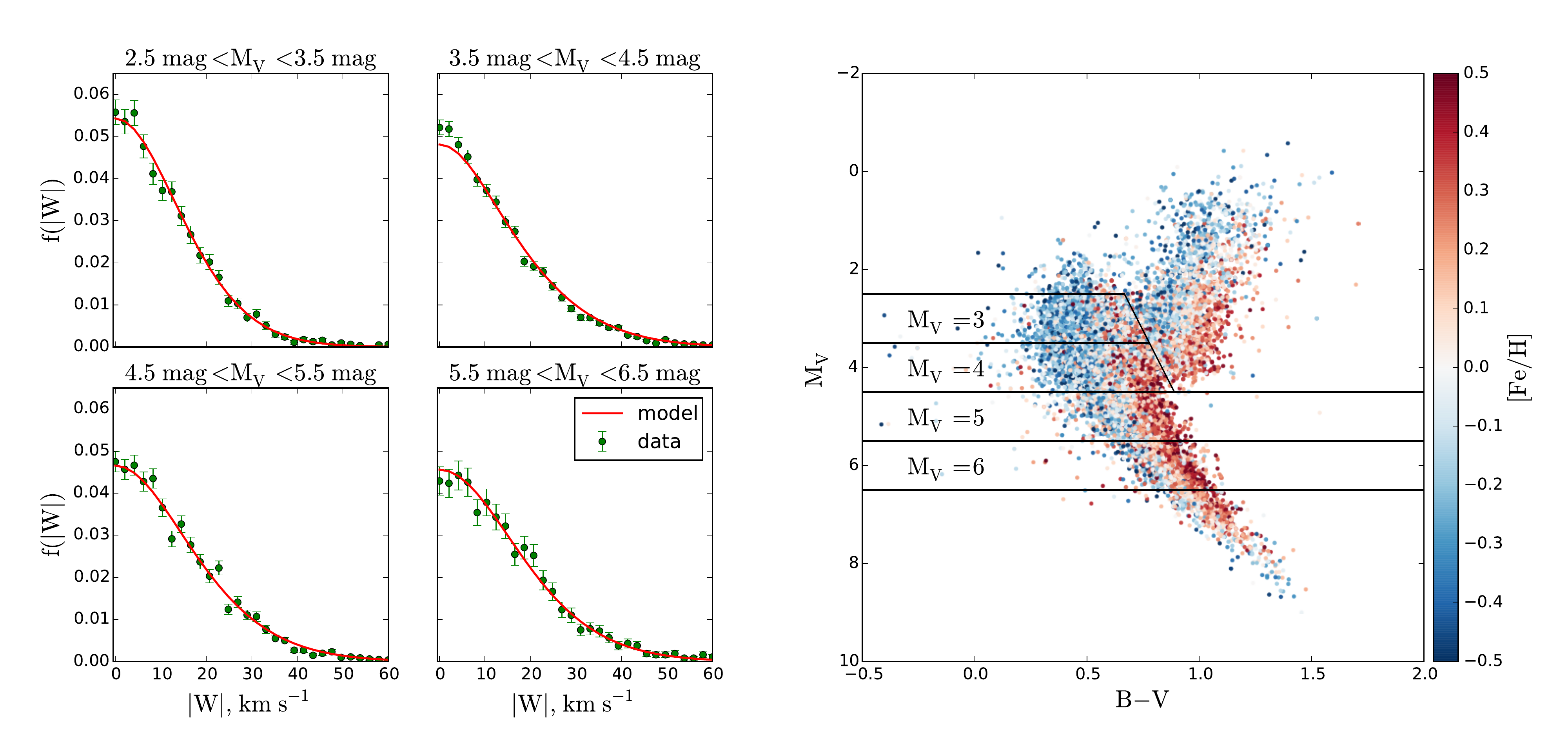}}
\caption{Normalised velocity distribution functions $f(|W|)$ for the different $M_{V}$ cuts. The selection of the MS stars and separation into magnitude bins
is shown on the right, where the CMD of the full TGAS$\times$RAVE thin-disc sample is plotted with APASS photometry and the RAVE metallicity is colour-coded.}
\label{fWMv_plot}
\end{figure*}

The first and most straightforward test that can be conducted with the JJ model is a comparison of the observed and predicted vertical density profiles  
because the primary focus of the model is the vertical structure of the stellar populations in the disc. 
By following the steps described in Sections \ref{table}-\ref{verted}, we calculated the vertical stellar density law for all stars in the cylinder, 
$n(\tilde{z}) = N^V_{\tilde{z}}/V_{\tilde{z}}$ with $V_{\tilde{z}}$ being the volume of the corresponding vertical bin.  
The top panel of Fig. \ref{Nz_plot} shows the stellar number density as calculated from the data (blue line) and as predicted by the model (violet and red curves).
The final model prediction is plotted in red, while the violet curve is given for comparison to show the number density law as predicted without 
the impact of the vertical distance error taken into account. The procedure of weighting the model output with the PDF 
$P(\tilde{z}|\sigma_z,z)$ results in migration of stars to larger observed heights $\tilde{z}$, which in turn leads to 
a better consistency with the data far from the midplane. On the other hand, the effect is very small close to the Galactic plane, predictably in view of 
the negligible distance error in close vicinity to the Sun (Fig. \ref{verted_plot}, top panel). The observed and the modelled density profiles, 
agree well at all heights, which also implies also a good agreement in the local region close to the Galactic plane where the model is well-calibrated and is therefore expected to produce  
robust predictions. By analogy to the vertical profiles, we derived a distribution of stars over the observed heliocentric distances that allows us to quantify the model-to-data agreement
in spherical volumes as well. The error in the predicted stellar number within a 25 pc local sphere (considered for the model calibration in \citetalias{rybizki15})
is $\sim$6\%. However, we note that the actual number of observed stars in this volume is smaller than ten, such that the 
statistics is strongly influenced by Poisson noise.

%The difference reaches 10\% for 200 pc sphere, the volume investigated with the JJ-model in Paper I and Paper III. 
When investigated in terms of the cumulative number of stars as a function of observed height, the model and data demonstrate 7.8\% disagreement at $\tilde{z}=1$ kpc
(Fig. \ref{Nz_plot}, middle panel). The ratio of the observed to modelled stellar number in individual vertical bins as shown
in the bottom panel of Fig. \ref{Nz_plot} helps to understand the systematic trends in the model-to-data differences. The ratio is close to unity
over a wide range of heights, which means that the robustness of the modelled vertical profile and our treatment of the stars that are missing in the volume as a result of the lack of 
chemical abundances and low quality of other parameters are reliable (Fig. \ref{missing_stars}). It is also clear that starting from $\sim$600 pc, the Poisson noise begins
to dominate star count statistics in the bins. From a height of approximately 800 pc, the number of stars is systematically under-represented in the model. 
Consistently, we expect this region to be biased in our simulation as we did not model stars at heights larger than 1 kpc and thus did not account 
for their possible presence in our sample owing to the large distance errors.

It is also interesting to perform a comparison in the space of observed parallaxes in order to validate our two-step treatment of the parallax uncertainties and the vertical distance errors.
Following our standard steps, we calculated the parallax distributions at individual model heights z 
and then weighted the results with the corresponding probabilities $P(\tilde{z}|\sigma_z,z)$ 
and completeness factors $S_Q(\tilde{z})$. The final distribution is shown in Fig. \ref{Par_plot}. Both data and model are normalised to the total number of observed and predicted stars, 
such that the shapes of the distributions can be clearly compared. We find a very good agreement between the two curves which demonstrates the reliability of our forward-modelling approach. 

\subsection{Vertical kinematics}\label{fW_results}

It is illustrative to test the model performance in terms of the vertical kinematics that was originally used to calibrate the model parameters \citepalias{just10}. 
Again, following the routine described in Sections \ref{table}-\ref{verted}, we evaluated for each observed height $\tilde{z}$ the velocity distribution function $f(|W|)$ 
as a superposition of Gaussians with the standard deviations given by the AVR. 
We considered the absolute W-velocities in a range of $0...60$ km s$^{-1}$ with a step of $\Delta |W| = 2$ km s$^{-1}$. 
The width of the step was selected such that it allowed tracing the shape 
of the velocity distribution function but at the same time was larger than the typical observational error in the velocity bin.
Following Eq. \ref{Qz2}, we modelled $f(|W|)$ in six horizontal slices probing the dynamical heating of the disc as a function of distance from the midplane.
The observed and predicted normalised velocity distribution functions calculated for the different ranges in $\tilde{z}$ as well as for the whole cylinder are shown in Fig. \ref{fWz_plot}. 
Green points represent the data with the error bars calculated as standard deviation of the Poisson distribution. The red curves are the model predictions.
In general, the observed and modelled vertical kinematics of the sample agree very well. 
The total predicted $f(|W|)$ (Fig. \ref{fWz_plot}, right plot) is consistent with the data within 1$\sigma$ at almost all $|W|$. This is also true for 
the velocity distribution functions compared to the data at different heights (Fig. \ref{fWz_plot}, three left columns). A noticeable discrepancy appears in the two lowest bins, 
for 0 pc $<\tilde{z}<$ 100 pc and 100 pc $<\tilde{z}<$ 200 pc where the fraction of the dynamically cold populations is underestimated in our model. 
In general, a trend with height is clearly evident: with the increase of the distance to the midplane, the shape of $f(|W|)$ becomes less peaked at small $|W|$. 
This is a natural and straightforward result as more dynamically heated populations reach larger distances from the Galactic plane because their velocities are higher.
\begin{figure*}
\centerline{\includegraphics[scale=0.5]{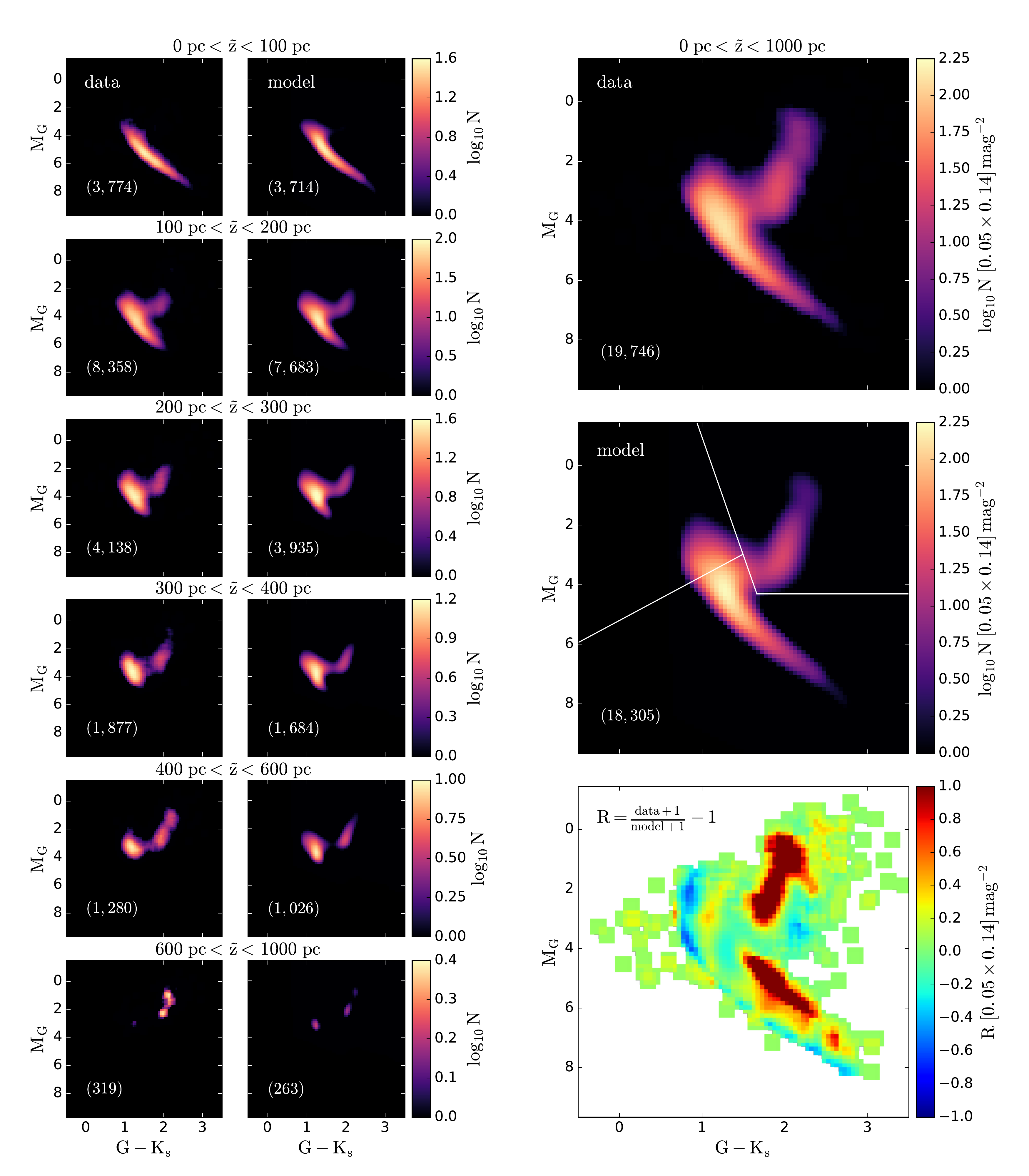}} % 0.5
%\centerline{\resizebox{\hsize}{!}{\includegraphics{All_hess.pdf}}}
\caption{\textit{Left.} Hess diagrams in $(M_G,G-K_s)$ as derived from the data and predicted by the model (left to right) 
shown for the different horizontal slices. The height increases from top to bottom. The numbers of stars are given in brackets.
\textit{Right.} Total Hess diagrams (top and middle) and their relative ratio (bottom). The white lines in the middle plot 
define the separation into the UMS, LMS, and RGB regions.}
\label{hess_plot}
\end{figure*}

We also compared $f(|W|)$ in four magnitude bins, for which we selected different parts of the MS (CMD in Fig. \ref{fWMv_plot}). The bin ranges were
the same as used in our previous work (compare to Fig. 3 in \citetalias{just10}). The CMD shows the whole 
TGAS$\times$RAVE thin-disc sample colour-coded with the RAVE metallicity. A metallicity gradient 
across the MS is visible such that each of the defined magnitude bins contains a mixture of stars with different metallicities and ages. 
To model $f(|W|)$ for the selected magnitude bins, we added the same colour-magnitude cuts to the calculation procedure and removed the stars that did not 
fall under the given criteria after adding the reddening, at the stage of applying the TGAS$\times$RAVE selection function (Section \ref{incompl}). 
The remaining calculation procedure was the same. 
The velocity distribution functions for the MS also show a good consistency with the data (Fig. \ref{fWMv_plot}, left), although 
the model slightly underestimates the role of the dynamically cold populations for the bin $3.5 \, \mathrm{mag}<M_V<4.5 \, \mathrm{mag}$.

\subsection{Hess diagrams}\label{hess_results}

\begin{figure}[ht!]
\centerline{\resizebox{\hsize}{!}{\includegraphics{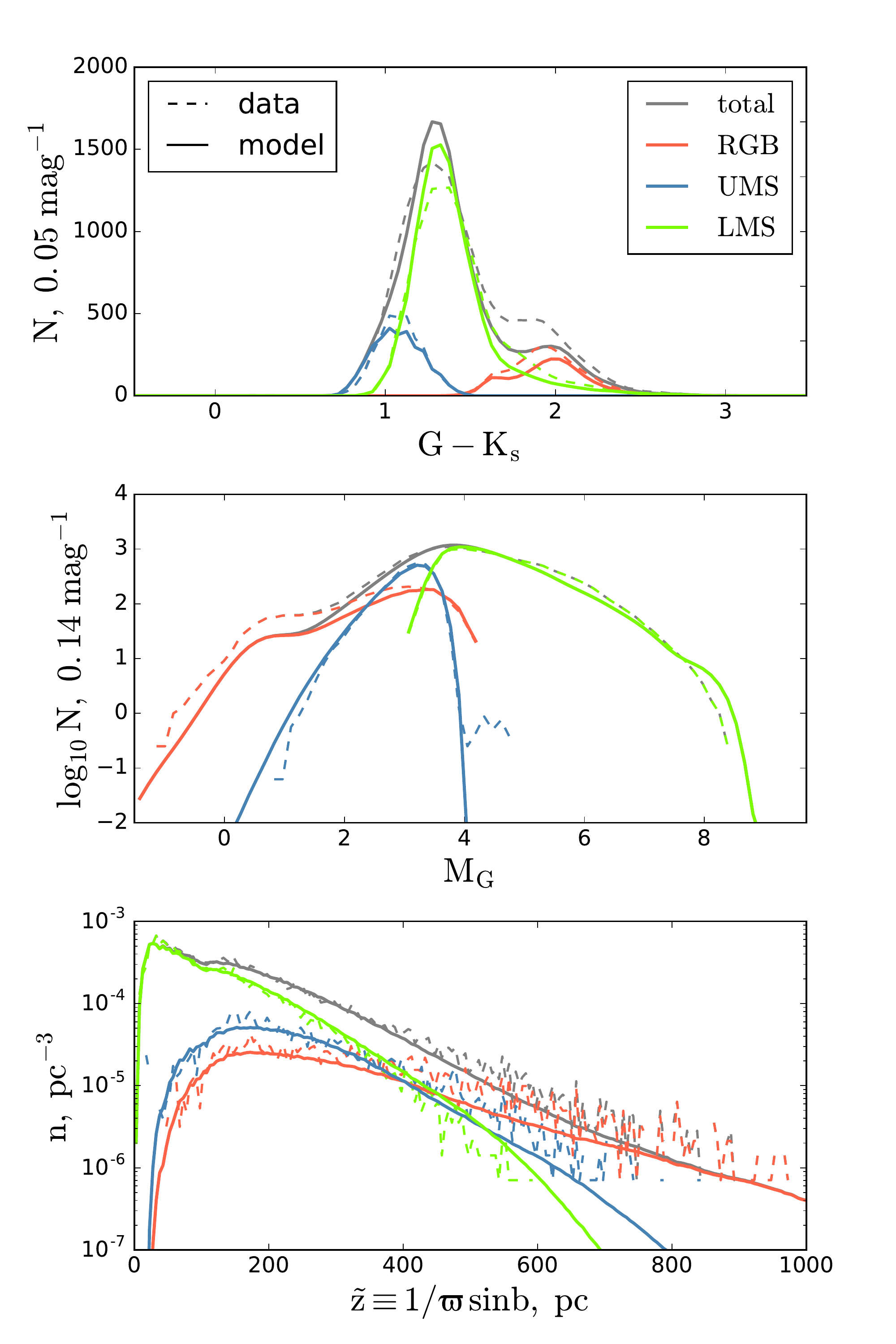}}}
\caption{UMS, LMS, and RGB stars studied separately.
\textit{Top.} Colour distributions as observed and modelled for the whole local cylinder.
\textit{Middle.} Luminosity functions for the same three populations. 
\textit{Bottom.} Stellar number densities as functions of observed height. The grey line corresponds to the red curve in Fig. \ref{Nz_plot}.}
\label{3pop_plot}
\end{figure}

In order to achieve deeper insight into the space distribution of the thin-disc populations, we also investigated the modelled 
cylindric volume in the 2D colour-magnitude space, that is, we built Hess diagrams. We used the Gaia $G$ and 2MASS $K_s$ filters and constructed the Hess diagrams 
in $(G-K_s,M_G)$ within the range of colours and absolute magnitudes of $-0.5...3.5$ mag and $9.5..1.5$ mag, respectively,
with corresponding steps of $\Delta (G-K_s)=0.05$ mag and $\Delta M_G=0.14$ mag. The typical uncertainty of the visual 
G-band magnitude is $\sim$0.003 mag and for the colour $(G-K_s)$, it is larger by about a factor of ten, $\sim$0.025 mag. 
Both values are well within our resolution on the Hess diagram axes, thus the choice of $G$ and $K_s$ photometry 
allows us to construct high-quality Hess diagrams without the need to complicate the simulations by adding photometric errors. 

The calculation procedure in this case differs from the one described previously at the stage of accounting for the vertical distance error.
For each modelled height $z$, we visualised the stellar content of this thin horizontal slice on the colour-absolute magnitude axes.
The absolute magnitude is itself a logarithmic function of the distance related to $z$ through $d(z)$ (Fig. \ref{verted_plot}, top panel). 
This additional distance dependence is the reason why we cannot model the resulting Hess diagram by appropriate weighting of 
the predictions calculated at the true heights in accordance to Eqs. \ref{Qz1} and \ref{Qz2}, as we did for the other quantities. 
In order to account for the impact of the 
populations lying at some true height $z_1$ on the Hess diagram modelled for the height $z_2$, 
we took into account that the populations from $z_1$, when observed at $z_2$, has different absolute magnitudes shifted by $\Delta M = 5 \log{(z_1/z_2)}$. 
Thus, the vertical error effect adds a spread on the vertical axis on the Hess diagram. 

To approach the problem, we rewrote Eq. \ref{Pz} in terms of the absolute magnitudes:
\begin{eqnarray}
&& P(\tilde{M}|\sigma_M,M) = \\ \nonumber 
&& = \frac{1}{C} \exp{\Big[ - \frac{1}{2}\Big(\frac{5}{\ln{10} \sigma_M} \Big)^2 \Big(10^{0.2(\tilde{M}-M)} - 1 \Big)^2 \Big]} \\ \nonumber 
&& \mathrm{with} \ C = \int_{-\infty}^{\infty} P(\tilde{M}|\sigma_M,M) d \tilde{M}.
\label{magpdf}
\end{eqnarray}
Here $P(\tilde{M}|\sigma_M,M)$ gives the probability of a star with the true absolute magnitude $M$ to be associated with another magnitude $\tilde{M}$
given a magnitude error $\sigma_{M}$. 
\ifx
The limits of integration $M_{min}$ and $M_{max}$ formally correspond to $\mp \infty$, but we 
replace them with the values of the physical meaning. We put $M_{min}=-8$ mag and $M_{max}=17$ mag as these are the maximum and minimum 
values of the absolute $G$-magnitudes as predicted by our set of the MIST isochrones. 
\fi
The error in the absolute magnitude is a function of distance, that is, of the height above or below the midplane: 
\begin{equation}
\sigma_M = \frac{5}{\ln{10} d(z)} \sigma_d(z)
\label{sigM}
\end{equation}
Now, we can smooth the Hess diagrams predicted for the different true heights $z$ by taking the corresponding magnitude errors $\sigma_{M}(z)$
from Eq. \ref{sigM}. We refer to the value in some row of the Hess diagram calculated for a given $z$ as $H_M$. Then the smoothing procedure can be expressed as
\begin{equation}
\tilde{H}_{M} = \sum_{M=M_1}^{M=M_2} H_{M} P(\tilde{M}|\sigma_M,M).
\label{HM1}
\end{equation}
The limits $M_1$ and $M_2$ are 
related to the range of the heights of interest $\tilde{z}_1 < \tilde{z} < \tilde{z}_2$:
\begin{eqnarray}
M_1 = M + 5 \log{(d(z)/d(\tilde{z}_1))} \\ \nonumber
M_2 = M + 5 \log{(d(z)/d(\tilde{z}_2))}
\label{HM2}
\end{eqnarray}
%We also check that the values of the limits $M_1$ and $M_2$ lay within our range of modelled magnitudes, 
%e.g., when the lower limit $M_1$ turns into $-\infty$ at $\tilde{z}_1=0$ kpc we replace it with the value of 10 mag. 
The Hess diagram $\tilde{H}$ smoothed in such a way shows where in the absolute magnitude-colour plane 
the populations from the height $z$ will appear when the range of observed heights $|\tilde{z}_2-\tilde{z}_1|$ is considered. 
The resulting Hess diagram for the given range of heights can be evaluated as a sum over all $\tilde{H}$:
\begin{equation}
H_{\tilde{z_1} < \tilde{z} < \tilde{z_2}} = \sum_i S_{Q}(z_i) \tilde{H}_i. 
\label{HM3}
\end{equation}

We constructed Hess diagrams for the same six vertical bins as were used in Section \ref{fW_results}. 
Observed and modelled Hess diagrams were additionally smoothed with a window $0.1\times0.28$ mag$^2$ (2$\times$2 bins in $(G-K_s,M_G)$). 
The results are plotted in Fig. \ref{hess_plot}. The two left columns show the Hess diagrams as derived from the data and predicted by the model 
given for increasing $\tilde{z}$ (top to bottom). A pronounced vertical trend appears: close to the Galactic plane, 
the LMS stars dominate; at the middle heights, the contribution of the the UMS becomes important and 
RGB stars appear as well; and starting from $\tilde{z} \approx 400$ pc, 
the stellar populations in Hess diagrams are represented by UMS and RGB alone. All these changes are very confidently traced in the model. 
The right column in Fig. \ref{hess_plot} shows three plots for the whole modelled cylinder (top to bottom) corresponding to the observed and predicted Hess diagrams, 
as well as their relative ratio calculated with the suppression of noise in the almost empty bins. The relative ratio plot 
demonstrates a general good agreement between data and model as it has already been inferred from investigating the vertical number density profiles, 
although a comparison in 2D space reveals the sources of model-to-data differences in more details.   
The typical model-to-data deviations over the Hess diagram are found to be $\pm$20\%, with the red and blue regions indicating the problematic areas 
(see Section \ref{discus}). 

We further investigated the UMS, LMS, and RGB populations separately. We defined them with the white lines in the middle plot of the right panel of Fig. \ref{hess_plot}. 
The border between the lower and upper parts of the MS approximately corresponds to the stellar masses of 1.5 $M_{\odot}$. We plot 
two projections of the total Hess diagrams onto the magnitude and colour axis. The resulting colour distributions and luminosity functions are shown 
in the top and middle panels of Fig. \ref{3pop_plot}.
In addition to the general difference in the number of stars, we see more clearly that the modelled LMS is more peaked than the observed one, 
while the UMS and RGB regions are less pronounced in the model. The same information can be obtained from the individual vertical density laws 
plotted for the three populations (Fig. \ref{3pop_plot}, bottom panel). 
The LMS, UMS, and RGB regions are under-populated in the model by 3.6\%, 6\%, and 34.7\%, respectively.

\section{Discussion}\label{discus}

In this section we discuss several potential problems of our modelling procedure.

When we analyzed the Hess diagrams (Fig. \ref{hess_plot}), we found a few tenth of stars in the data that were clearly outliers 
(see the reddest and bluest $G-K_s$ values).
These might be misidentified stars as well as contamination of the metal-poor thick-disc and halo stars and objects with unrealistic 
metallicities, magnitudes, or colours. 
One of the problematic red areas is associated with the LMS region: the observed MS is noticeably wider than the predicted one. 
This might be caused by the impact of binarity, which is ignored in our modelling. Additionally, metal-rich stars are present in the 
data sample ($\mathrm{[Fe/H]}>0.2$), and their number is underestimated with the model AMR (see Fig.16 in \mbox{\citetalias{just10}}, 
the metallicity prescribed to the youngest stellar population is $+0.02$ dex). 
Moreover, we predict an underestimated stellar density in the RGB region. This we attribute to the simplicity of accounting for the $S_Q$ completeness factor,
which is included in the modelling as a function of height above the plane (Fig. \ref{missing_stars}, top panel), 
although it also shows a variation with colour and absolute magnitude (Fig. \ref{missing_stars}, bottom panel).
By ignoring this dependence of $S_Q$ on magnitude and colour, we may over- or underestimate stellar numbers 
in the regions of the Hess diagram where the values of incompleteness in Fig. \ref{missing_stars} deviate considerably from the average. 
This effect can also be responsible for a blue region near the UMS at $G-K_s \approx 0.8$ mag (compare to the bottom panel of \mbox{Fig. \ref{missing_stars}}).
On the other hand, when considered together with a blue region in the LMS (see colour range $1.8 \lesssim (G-K_s)/\mathrm{mag} \lesssim 2.8$ at the relative ratio plot in Fig. \ref{hess_plot}), 
it points to a small colour shift between the modelled and observed Hess diagrams of $\sim$0.1 mag in $G-K_s$.
Several
reasons may be responsible for this: an underestimated reddening, a systematic shift of isochrones, or a lack of metal-poor thin-disc populations in the data 
due to an incorrect separation of thin- and thick-disc stars;
%, or systematic errors in metallicity}
the interplay of all of these factors is also possible. Taking into account the locality of our volume, however, 
it is improbable that this shift is related to the underestimated reddening.  
We return to this question in view of the new 
results obtained on the basis of Gaia DR2 (see below). Furthermore, we discuss the sensitivity of our results to the choice of dust map and stellar library. 

\begin{figure*}[t!]
\centering
\includegraphics[scale=0.45]{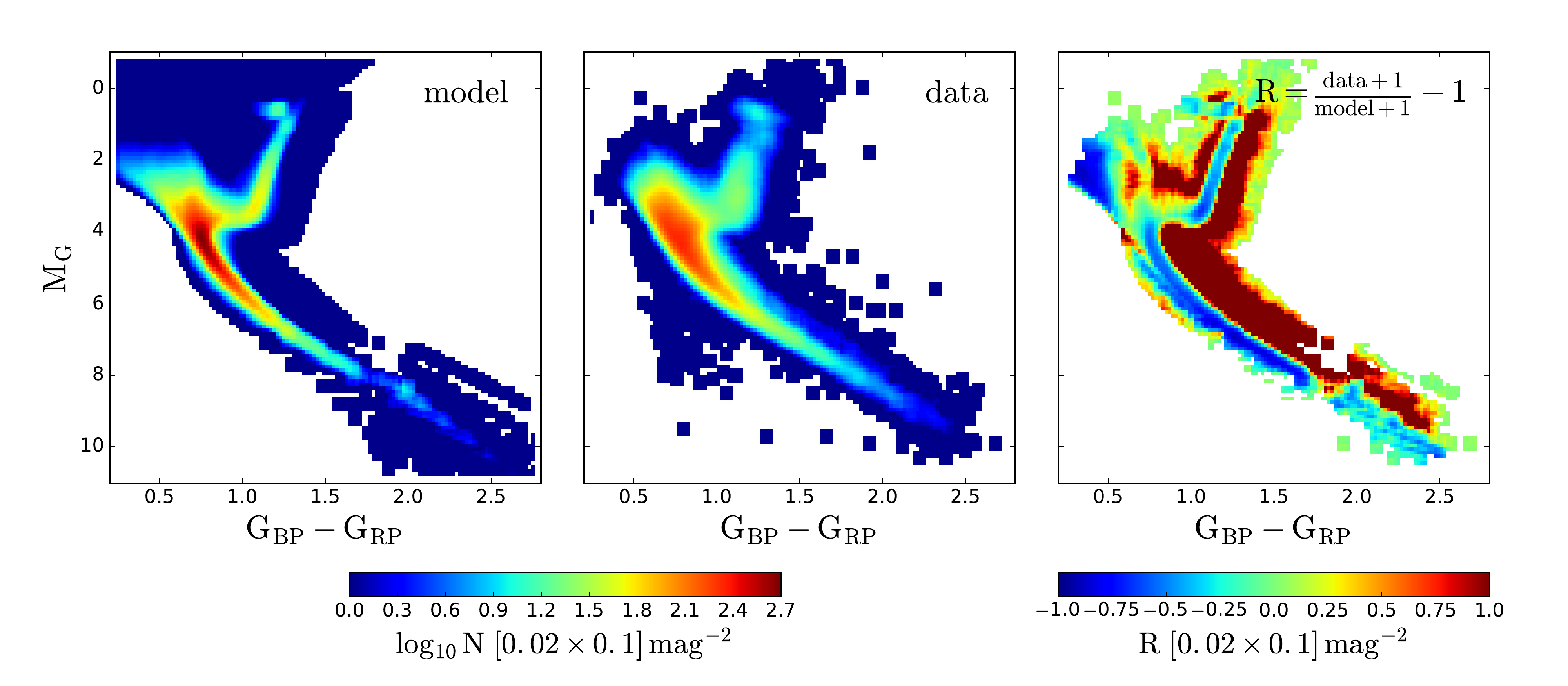}
\caption{Modelled (left) and observed (middle) absolute Hess diagrams of the the stars in common between RAVE and Gaia DR2. 
The right panel displays their relative ratio. 
The sample belongs to the solar cylinder of 300 pc radius and 1 kpc height 
below the midplane with stars selected in the sky area of the TGAS$\times$RAVE
sample (see Section \ref{cuts} and Fig. \ref{data_sample}). The thick disc is 
included in the modelling. Model and data are both smoothed with a
window of 0.04$\times$0.2 mag$^2$ (2$\times$2 bins in ($G_{BP}-G_{RP},M_G$)).}
\label{hess_abs_md}
\end{figure*}

A possible unreliability of the dust map can have a twofold impact on our results. 
First, the predicted Hess diagram may appear shifted in the colour-magnitude plane relative to the observed one. Second, as the data selection function strongly
depends on apparent magnitude and colour and the predicted number of stars may change significantly over the colour-magnitude bins 
(this is particularly related to the UMS and RGB regions), under- or overestimation of extinction and reddening may add up to the uncertainty 
in star counts.
The 3D dust map used in our simulations is mainly based on the high-angular resolution extinction model from \mbox{\citet{green15}} 
constructed by the statistical technique from photometric data of Pan-STARRS and 2MASS. The safe range of distances covered by this map 
lies approximately in a range from 300 pc to 4-5 kpc, although the values of the minimum and maximum reliable distance may significantly vary over the sky. 
In most of our modelled sky region, the minimum reliable distance is $\sim$200 pc and extends to 300 pc.  
This implies that reddening may not be modelled reliably in those regions of the cylinder that are in close vicinity to the Sun.
However, we do not expect the roughness of the extinction map in the nearby areas to have a significant effect on the modelling process 
because with the cut of $|b|>20^\circ$, we avoid the most problematic near-plane regions. 
To quantify the impact of the extinction model on our results, we tested an additional 3D reddening map from \mbox{\citet{capitanio17}}\footnote{\url{http://stilism.obspm.fr}}
when we used Gaia DR2 data (see below).  

With the full stellar evolution now included in the modelling, our predictions are sensitive to the choice of the stellar evolution package. 
To quantify the corresponding uncertainty, we tested an alternative set of isochrones generated by the 
PAdova and TRieste Stellar Evolution Code (PARSEC, \mbox{\citealp{bressan12}}, \mbox{\citealp{marigo17}}) for the same range of metallicities 
and ages as described in Section \ref{table}. The total number of predicted stars increased by $\sim$10\%, but no appreciable difference 
of the model behaviour was discovered apart from this. 

\begin{figure}[t!]
\centering
\includegraphics[scale=0.4]{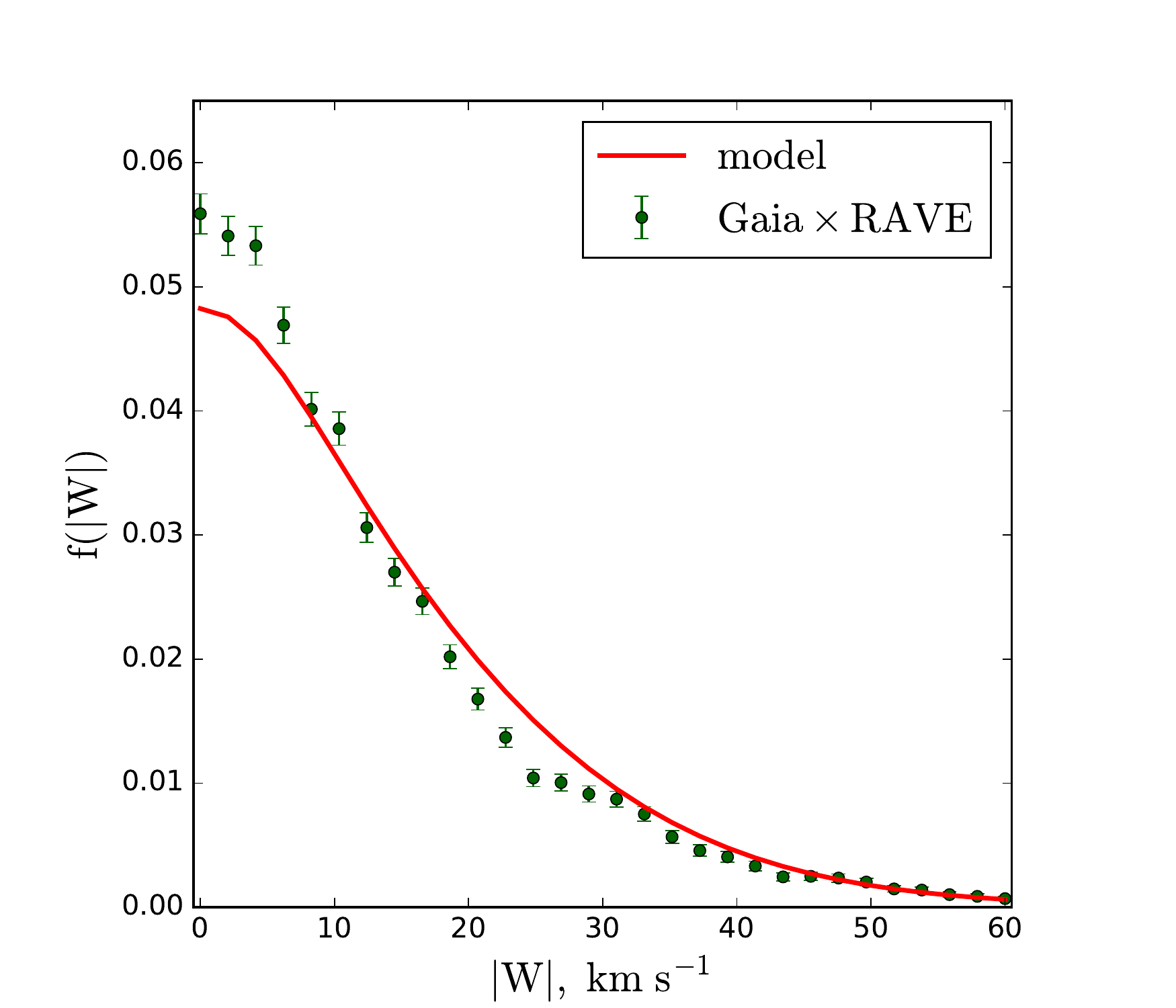}
\caption{Normalised $|W|$-velocity distribution function at $|z|<100$ pc. The data sample is based on Gaia DR2.}
\label{fw_local}
\end{figure}

As the Gaia DR2 astrometry and completeness are significantly improved with respect to TGAS
\mbox{\citep{lindegren18,arenou18}} and the Gaia radial velocities have an even better
precision than those of RAVE \mbox{\citep{katz18}}, we performed a sanity check complementary to the local test
using the Gaia DR2 astrometric parameters and radial velocities. 
To mimic the test presented in this paper, we used the Gaia$\times$RAVE cross-match with the Gaia radial velocities \citep{soubiran18}
and selected the local subset belonging to the same spatial volume and area on the sky as the TGAS$\times$RAVE sample
described in Section \ref{data}.
We did not attempt to repeat the full forward-modelling described in Section \ref{modelling} by accounting for parallax errors; instead,
we used geometric distances 
from \mbox{\citet{bailer-jones18}} derived in the framework of Bayesian approach on the basis of Gaia DR2 parallaxes. 
In this test we did not apply metallicity scatter to the AMR eiter, 
which allowed us to reduce the number of stellar assemblies and speed up the calculations. 
These simplifications do not influence the star counts much: (1) even if the distance errors reach $\sim$50\% 
for the individual stars at ~1 kpc, the overall stellar density at this heliocentric distance is quite low in our geometry, and the corresponding 
uncertainty introduced in the number of stars in the volume is not significant; 
and (2) the metallicity grid affects the modelled number of stars only indirectly through the sample selection function, which is sensitive to the 
apparent magnitudes and colours.
In order to avoid additional incompleteness, no cuts on RAVE Fe or Mg abundances were used. 
Correspondingly, the thick disc was added to the model as a mono-age population with a metallicity of $-0.7$ dex and a Gaussian dispersion of 0.2 dex. 
As some stars of the RAVE DR5 are missing in the Gaia$\times$RAVE cross-match (because of unknown Gaia radial velocities or match problems; 
this makes up $\sim$ 11\% of the full RAVE DR5), we built a modified RAVE selection function that incorporates this additional incompleteness and used it for our test.
The selected Gaia$\times$RAVE local sample contains 63,408 stars. We built absolute and apparent Hess 
diagrams using the updated Gaia DR2 $G$-band and $G_{BP}-G_{RP}$ colour from PARSEC isochrones. 
Because the metallicity scatter and distance errors were not modelled, all features of the predicted Hess diagram look 
more pronounced than those seen in the data (\mbox{Fig. \ref{hess_abs_md}}). 
Regardless of the width of the observed and modelled MS, we expect their centres to coincide; however, the right panel of Fig. \ref{hess_abs_md} shows that 
this is not the case, which implies a small colour shift of about $\sim$0.05 mag in $G_{BP}-G_{RP}$ between the data and the model. As the thick disc is included at this stage, 
the shift is most likely related to a systematic shift of the isochrones. 
We also observe an excess of stars in the UMS region which 
indicates that a better understanding of the Gaia selection function is necessary.
Additionally, there is a deficit of stars in the modelled LMS in the colour range of $1.7 \lesssim (G_{BP}-G_{RP})/\mathrm{mag} \lesssim 1.9$. 
As we did not include any special colour cuts and the applied selection function is essentially independent of colour, this 
feature must arise from the stellar library and needs to be further investigated in future work.
With this setup we reproduced star counts with an accuracy of $\sim$3\%.
We report the presence of non-reproduced dynamically cold populations at $|z|<100$ pc even more confidently than in
case of the TGAS$\times$RAVE test (Fig. \ref{fw_local}, also see the upper left panel of Fig. \ref{fWz_plot}). 
Finally, we compared the outcome of the sanity check star counts for the \mbox{\citet{bovy15}} and \mbox{\citet{capitanio17}} dust maps. 
We find only a difference between the two runs of $\sim$1\%. 
Thus, the outcome of our tests gives us a good insight into potential problems of the model and can be viewed as 
a first step to the future and more comprehensive work with Gaia DR2.

\section{Conclusions}\label{final}

We used the forward-modelling technique to test the semi-analytic JJ model. 
We selected a clean sample of thin-disc stars from the TGAS$\times$RAVE cross-match providing precise 
astrometry as well as radial velocities and chemical abundances.   
We investigated the large local volume of the solar cylinder with radius of 300 pc extending to 1 kpc below the Galactic plane. 
The full stellar evolution in the form of MIST stellar isochrones was implemented in the model. 
For each modelled height $z$, we accounted for the parallax errors and reddening, and reproduced the incompleteness of the sample. 
The results were assigned with the additional weights to model the effect of the distance error in the vertical direction. 

We found that the model predictions and data agree well with each other. A deviation of $\sim 1 \sigma$ is found between the 
modelled and observed velocity distribution functions $f(|W|)$ close to the Galactic plane, indicating that the role of the cold young stellar populations
may be underestimated in our local disc model. When complemented with the velocity distribution functions for the different magnitude bins of the MS stars, 
this implies that the source of the discrepancy in vertical kinematics is related to the lower part of the UMS region. 
A complementary test based on the RAVE and new Gaia DR2 data confirmed the results presented in Section \ref{results},
and the reported underestimation of dynamically cold populations is even more prominent when tested with Gaia DR2 astrometry and radial velocities. 
  
With this realistic performance test, we demonstrated the robustness of the local JJ model when compared to the data up to 1 kpc away 
from the Galactic plane. However, non-negligible deviations from the data are identified in case of the vertical kinematics of the disc, which suggests 
that the model should be adapted to match the new Gaia data. 
The next step will include an extension of the model to other Galactocentric distances: 
the code Chempy will be used to build the chemical evolution model of the disc, consistent with the adopted SFR and AVR 
varying with Galactic radius. 
For this purpose, Gaia DR2 and its next releases as well as high-resolution spectroscopic surveys such as APOGEE, Gaia-ESO, and GALAH will be of high benefit.

%\newpage
\section*{Acknowledgements}

This work was supported by Sonderforschungsbereich SFB 881 `The Milky Way
System' (subprojects A6 and A5) of the German Research Foundation (DFG).

Funding for RAVE has been provided by the Australian Astronomical
Observatory, the Leibniz-Institut f\"{u}r Astrophysik Potsdam (AIP), the
Australian National University, the Australian Research Council, the
French National Research Agency, the German Research Foundation (SPP 1177 and SFB 881),
the European Research Council (ERC-StG 240271 Galactica), the Istituto Nazionale di
Astrofisica at Padova, The Johns Hopkins University, the National Science Foundation of
the USA (AST-0908326), the W. M. Keck foundation, the Macquarie University, the
Netherlands Research School for Astronomy, the Natural Sciences and Engineering Research
Council of Canada, the Slovenian Research Agency, the Swiss National Science Foundation,
the Science \& Technology Facilities Council of the UK, Opticon, the Strasbourg Observatory,
and the Universities of Groningen, Heidelberg, and Sydney. The RAVE web site is at
\url{http://www.rave-survey.org}.

This work has made use of data from the European Space Agency (ESA)
mission {\it Gaia} (\url{https://www.cosmos.esa.int/gaia}), processed by
the {\it Gaia} Data Processing and Analysis Consortium (DPAC,
\url{https://www.cosmos.esa.int/web/gaia/dpac/consortium}). Funding
for the DPAC has been provided by national institutions, in particular
the institutions participating in the {\it Gaia} Multilateral Agreement.

%%%%%%%%%%%%%%%%%%%%%%%%%%%%%%%%%%%%%%%%%%%%%%%%%%%%%%%%%%%%%%%%%%%%%
%%%%%%%%%%%%%%%%%%%%%%%%%%%%%%%%%%%%%%%%%%%%%%%%%%%%%%%%%%%%%%%%%%%%%
\bibliographystyle{./aa.bst}
\bibliography{ld.bib}

\end{document}